\documentclass[11pt,a4paper]{article}
\pdfoutput=1 

\usepackage{jheppub} 

\usepackage[utf8]{inputenc}

\usepackage{mathtools}
\usepackage{graphicx}
\usepackage{enumitem}
\usepackage{dsfont}

\title{Superspace invariants and 3-point correlators in 3d $\mathcal{N}=3,4$ SCFTs}

 \author[a]{Aditya Jain}
 \author[b]{and Amin A. Nizami}

\affiliation[a]{Asia Pacific Center for Theoretical Physics (APCTP),\\Pohang, Gyeongsangbuk 37673, Korea}
\affiliation[b]{Department of Physics, Ashoka University,\\Rajiv Gandhi Education City, Rai, NCR 131029, India}

\emailAdd{adityaj2807@gmail.com}
\emailAdd{amin.nizami@ashoka.edu.in}

\newcommand{\cN}{\mathcal{N}}
\newcommand{\cO}{\mathcal{O}}
\newcommand{\cJ}{\mathcal{J}}

\newcommand{\A}{\alpha}
\newcommand{\B}{\beta}
\newcommand{\M}{\mu}

\newcommand{\LL}{\lambda}

\abstract{We use the auxiliary polarization spinor formalism together with superspace techniques to construct a complete and minimal list of 3-point invariant structures in three-dimensional superconformal field theories (SCFTs) with $\cN=3$ and $\cN=4$ superconformal symmetry. The existence of non-abelian $R$-symmetry for $\cN=3,4$ gives rise to novel invariant structures built from the antisymmetric invariant tensor. These invariants are used to enumerate the structural form of spinning 3-point correlators of general as well as conserved spinning superfield operators in 3d SCFTs. For conserved operators, we find that the $\cN=3$ correlators are fixed upto one parity-even and one parity-odd structure, while $\cN=4$ conserved correlators admit two parity-even structures and one parity-odd structure, with the second parity-even structure associated with mirror symmetry breaking. 
}

\makeatletter
\g@addto@macro\bfseries{\boldmath}
\makeatother

\begin{document} 
\maketitle
\flushbottom

\section{Introduction}
\label{sec:intro}

Conformal symmetry together with supersymmetry imposes very strong constraints on quantum field theories. 
The local operators of a superconformal field theory (SCFT) organize themselves into supermultiplets, and kinematic and symmetry considerations fix 2- and 3-point functions essentially completely, up to 
scaling dimensions and OPE coefficients of primary operators. For spinning operators, 
even the kinematic problem is non-trivial: one must first construct a complete and 
minimal basis of superconformally invariant 3-point structures, and then construct structures for general 3-point functions using these invariants. Imposing 
conservation constraints (shortening conditions) on the superfield operators introduces extra restrictions on the structure of the correlators. 

The focus of this work will be the constraints on the form of 3-point correlators of superfield operators with spin in three dimensional superconformal field theories (SCFTs) with $\mathcal{N}=3, 4$ supersymmetry. This effectively completes the line of work initiated in \cite{AAN13} and developed further in \cite{AJ22} for 3d SCFTs. Like these earlier works, we use the superspace formalism for SCFT correlators as developed by Park and Osborn \cite{Park97, Park98, Park99, Osborn98}, together with polarization spinor techniques to encode spin as employed in \cite{GPY11} to enumerate three point invariants and correlators of spinning operators in 3d (non-supersymmetric) CFTs. Our results apply to 3-point functions of general (non-conserved) spinning operators with $\mathcal{N}=3, 4$ supersymmetry in 3d SCFTs, and include as special cases the case of conserved operators where shortening conditions on the supermultiplets impose additional constraints on the correlator.

The polarization spinor formalism of \cite{GPY11} established that parity-odd structures 
in 3-point functions of conserved currents in 3d CFTs are associated with interacting 
Chern-Simons-matter theories and are absent in free theories. This connection to 
parity-odd structures, together with the relevance of higher-spin symmetry 
\cite{MZ1,MZ2}, makes the systematic enumeration of 3-point invariants a problem of 
broad utility. The key idea of the program is that a general spinning 3-point SCFT 
correlator can be expressed in terms of kinematical superconformal invariant structures:
\begin{align}
\left\langle \cO_{s_{1}}\cO_{s_{2}}\cO_{s_{3}}\right\rangle
\sim
\lambda_{1}^{2s_{1}}\lambda_{2}^{2s_{2}}\lambda_{3}^{2s_{3}}
\Rightarrow
\sum\!\!\!
\begin{tabular}{c}\small
monomials of \\[-.2em]\small invariants
\end{tabular},
\label{eq:finalcorrelatorschematic}
\end{align}
where $\lambda$'s are commuting auxiliary spinors encoding the spin of the operators. 
For 3d $\cN=1$ SCFTs this program was initiated in \cite{AAN13}, and the non-trivial 
$R$-symmetry structures for $\cN=2$ --- the $V$-matrix and the associated fermionic 
invariants --- were handled in \cite{AJ22}, with free theory computations providing 
consistency checks. The $\cN=3,4$ case is of special interest because the $R$-symmetry 
group is non-abelian,\footnote{For $\cN=1$ there is no $R$-symmetry while for $\cN=2$ 
the $R$-symmetry group $SO(2)$ is abelian. See \cite{AAN13,AJ22}.} which leads to 
novel 3-point invariants built from the antisymmetric invariant tensor 
$\epsilon^{a_1\cdots a_\cN}$ of $SO(\cN)$, absent for $\cN\leq 2$. These 
$\epsilon$-invariants are responsible for the second parity-even structure in conserved 
$\cN=4$ correlators. We will enumerate all the invariants, both parity-even and parity-odd, as well as bosonic and fermionic, and construct 3-point functions of spinning operators using the invariants. The parity-odd structure in conserved correlators is found to obey 
triangle inequalities in the operator spins, extending the analogous result of 
\cite{BS23a} for $\cN=1$.

The structure of 3-point functions of conserved current multiplets in 3d $\cN$-extended 
SCFTs has been studied in 
\cite{BKS15,BKS15b,KS16}. 
In \cite{BKS15}, the 2- and 3-point functions of the supercurrent and flavour current 
multiplets were computed for $\cN\leq 3$: the supercurrent 3-point function is fixed 
by a single structure for all $\cN\leq 3$, while for flavour current multiplets one 
structure appears for $\cN=1,3$ and two for $\cN=2$. The $\cN=4$ case \cite{BKS15b} 
yields two independent supercurrent structures, one present in all $\cN=4$ SCFTs and 
one appearing only in theories not invariant under the mirror map 
$SU(2)_L\leftrightarrow SU(2)_R$; two inequivalent flavour current multiplets exist, 
each with a single tensor structure. For $\cN=5,6$ \cite{KS16} the supercurrent 
correlator is again uniquely fixed, consistent with every $\cN=5$ theory being 
mirror-symmetric as an $\cN=4$ theory.

A complementary program for $\cN=1$ in 3d was developed by Buchbinder and Stone 
\cite{BS21,BS21b,BS23a,BS23b}, establishing that Grassmann-even conserved 3-point 
functions are fixed by one parity-even and one parity-odd structure (with the parity-odd 
subject to spin triangle inequalities), while Grassmann-odd correlators admit a unique 
parity-even structure. The present work extends this to $\cN=3,4$. 

Our results for the 
supercurrent are consistent with \cite{BKS15,BKS15b}: conservation fixes the $\cN=3$ 
correlator to one parity-even and one parity-odd structure, while for $\cN=4$ two 
parity-even and one parity-odd structure survive, with the second parity-even structure 
arising from the $\epsilon$-invariants and associated with mirror symmetry breaking. However, the above works do not address general (non-conserved) spinning operator correlators in $\cN=3,4$ SCFTs.
We would like to emphasize that the techniques we employ enable us not only to reproduce the existing results for conserved correlators in $\cN=3,4$ 3d SCFTs and extend them to higher spin cases, but we also enumerate the tensor structures of 3-point correlators with spinning {\it non}-conserved operators.

This paper is organized as follows. Section~\ref{sec:methods} reviews the superspace 
and polarization spinor methods used throughout. 
In section~\ref{sec:invariants} the superspace invariants for $\mathcal{N}=3$ and $\mathcal{N}=4$ 3d SCFTs, including the novel $\epsilon$-invariants, are systematically constructed and relations between them delineated to provide a complete minimal list of invariants. 
In section~\ref{sec:correlators} the tensor structures of various 3-point correlators with spinning operators for $\mathcal{N}=3$ and $\mathcal{N}=4$ 3d SCFTs are enumerated using the invariants previously constructed for this purpose. This section also contains the constraints of conservation, when one or more operators saturate the unitarity bound, leading to additional constraints on the correlators.
Section~\ref{sec:discussion} concludes with a summary 
and future directions. Conventions and supplementary relations are collected in 
appendices~\ref{app:conventions} and~\ref{app:epsilonrels}.

\section{Superspace methods for 3d $\cN$-extended SCFTs}
\label{sec:methods}

The group-theoretic approach for studying correlation functions in conformal and superconformal theories was systematically developed in the 1990s by Osborn and collaborators \cite{Osborn94,Erd97,Dolan01,Park97,Park98,Park99,Osborn98}. In particular, Park's seminal work on the superspace formulation of $\cN$-extended SCFTs in $d=3,4,6$ \cite{Park97,Park98,Park99} serves as a central reference for modern studies of superconformal correlators and bootstrap constraints. Using superspace coordinates in tandem with infinitesimal and finite superconformal transformations, the Park-Osborn formalism determines a complete set of covariant building blocks sufficient to impose superconformal invariance on general 3-point correlators. In this section we review this construction, highlighting the role of the \textit{superinversion} transformation and the $V$-matrix, both of which will be essential for the systematic construction of superconformal invariants in Section~\ref{sec:invariants}.


\subsection{Superspace and superinversion}
\label{sec:superinv}

In 3d Minkowski spacetime, the bosonic part of the $\cN$-extended superconformal algebra is $o(2,3)\oplus o(\cN)$, where $o(\cN)$ represents the internal $R$-symmetry part of the superconformal group \cite{Park99}. The superspace for 3d $\cN$-extended SCFTs looks like
\begin{align}
z^{A}\equiv \{x^{\M}, \theta^{a\A}\},\quad
\begin{tabular}{l}
$\M=0,1,2,$\\
$\A=1,2,\quad a=1,2,\hdots,\cN$.
\end{tabular}
\end{align}
The coordinates $\theta^{a\alpha}$ are Majorana spinors and transform in the vector representation of the $R$-symmetry group $SO(\cN)$. 

The supersymmetry algebra in 3d has the standard form:
\begin{align}
\{Q^{a}_{\alpha},Q^{b}_{\beta}\}=2(\sigma^{\mu})_{\alpha\beta}P_{\mu},\quad 
[P_{\mu},P_{\nu}]=[P_{\mu},Q^{a\alpha}]=0,
\end{align}
where $Q^{a\alpha}$ are supersymmetry generators, that transform as Majorana spinors and generate translations in the $\theta$-coordinate. The use of superspace lets us define a \textit{supercovariant derivative},
\begin{align}
D^{a}_{\alpha}=\frac{\partial}{\partial \theta^{a\alpha}}+\frac{i}{2}\theta^{a\beta}(\sigma^{\mu})_{\beta\alpha}\partial_{\mu}.
\label{eq:supercovderiv3dNext}
\end{align}
This derivative operator satisfies
\begin{align}
\{D^{a}_{\alpha},Q^{b}_{\beta}\}=0,\qquad
\{D^{a}_{\alpha},D^{b}_{\beta}\}=-2(\sigma^{\mu})_{\alpha\beta}P_{\mu}.
\end{align}

For two superspace points $(z_{i}^{A},z_{j}^{B})$, the structures that are annihilated by the supersymmetry generators $Q^{a\alpha}$ are \cite{Osborn98,Park99,AAN13}
\begin{align}
\tilde x_{ij}^{\mu}=x_{ij}^{\mu}+\frac12 \theta_{i}^{a\alpha}(\sigma^{\mu})_{\alpha\beta}\theta_{j}^{a\beta},\qquad \theta_{ij}=\theta_{i}-\theta_{j},
\end{align}
where $\tilde x^{\mu}_{ij}$ is also called the supersymmetric interval. In spinor notation, one gets
\begin{align}
(\tilde{X}_{ij})_{\alpha}^{\ \beta}=(X_{ij})_{\alpha}^{\ \beta}+i\theta^a_{i\alpha}\theta_{j}^{a\beta}
+\frac{i}{2}(\theta^a_{i}\theta^a_{j})\delta_{\alpha}^{\ \beta}.
\label{eq:supersymmetric3dobjects}
\end{align}
While these objects can be used to construct super-Poincar\'e invariant structures, building superconformal invariants requires exploiting the relation $S\sim I\cdot Q\cdot I$, where $S$ is the special superconformal transformation and $I$ is the \textit{superinversion}, the supersymmetric analogue of conformal inversion. Under superinversion, the superspace coordinates transform as \cite{Osborn98,Park99,AAN13}
\begin{align}
x^{\mu}\to \frac{x^{\mu}}{x^{2}+\frac1{16}(\theta^{a}\theta^{a})^{2}},\qquad
\theta^{a}_{\alpha}=-(X_{+}^{-1}\theta^{a})_{\alpha},\quad
\theta^{a\alpha}\to -(\theta^{a}X_{-}^{-1})^{\alpha},
\label{eq:superinv3d}
\end{align}
where
\begin{align}
(X_{\pm})_{\alpha}^{\ \beta}=X_{\alpha}^{\ \beta}\pm \frac{i}{4}(\theta^{a}\theta^{a})\delta_{\alpha}^{\ \beta},\quad
X_{+}X_{-}=\bar x^{2}\mathds{1}, \quad
\bar x^{2}=x^{2}+\frac{1}{16}(\theta^{a}\theta^{a})^{2},
\label{eq:1ptblocks3d1}
\end{align}
with inverses
\begin{align}
X_{\pm}^{-1}=\frac{1}{\bar x^{2}}X_{\mp}.
\label{eq:1ptblocks3d2}
\end{align}
Under superinversion, $X_{\pm}$ transform as
\begin{align}
X_{\pm}\to X_{\pm}^{-1}.
\label{eq:1ptblocks3d3}
\end{align}
Spinor indices have been suppressed; conventions for raising, lowering, and contraction of spinor indices are given in appendix~\ref{app:conventions}.

\subsection{Covariant building blocks}
\label{sec:covbuilding}

Using $X_{i\pm}$ and $\theta_{i}$'s along with their superinversion transformations above, we can construct 2-point objects that transform homogeneously under superinversion.
\begin{gather}
(X_{ij+})_{\alpha}^{\ \beta}=(X_{i+})_{\alpha}^{\ \beta}-(X_{j-})_{\alpha}^{\ \beta}+i\theta_{i\alpha}^{a}\theta_{j}^{a\beta},\\
(X_{ij-})_{\alpha}^{\ \beta}=(X_{i-})_{\alpha}^{\ \beta}-(X_{j+})_{\alpha}^{\ \beta}-i\theta_{j\alpha}^{a}\theta_{i}^{a\beta},
\label{eq:2ptblocks3d}
\end{gather}
These are related to the supersymmetric interval by
\begin{align}
X_{ij\pm}=\tilde X_{ij}\pm \frac{i}{4}(\theta_{ij}^{a}\theta_{ij}^{a})\mathds{1},
\end{align}
and satisfy
\begin{align}
X_{ij+}X_{ij-}=\bar x_{ij}^{2}\mathds{1}, \qquad
\bar x_{ij}^{2}=\tilde x_{ij}^{2}+\frac{1}{16}(\theta_{ij}^{a}\theta_{ij}^{a})^{2},
\label{eq:2ptblocksscalar3d}
\end{align}
so that
\begin{align}
X_{ij\pm}^{-1}=\frac{1}{\bar x_{ij}^{2}}X_{ij\mp}.
\label{eq:2ptblocksinv3d}
\end{align}
Their superinversion transformations are
\begin{align}
\begin{gathered}
X_{ij+}\to X^{-1}_{i+} X_{ij+}X_{j-}^{-1},
\quad
X_{ij-}\to X^{-1}_{j+} X_{ij-}X_{i-}^{-1},\\
X_{ij+}^{-1}\to X_{j-} X^{-1}_{ij+}X_{i+},
\quad
X_{ij-}\to X_{i-} X^{-1}_{ij-}X_{j+},
\end{gathered}\qquad
\bar x_{ij}^{2}\to \frac{1}{\bar x_{i}^{2}\bar x_{j}^{2}}\bar x_{ij}^{2}.
\label{eq:2ptblocks3dsuperinv}
\end{align}
Systematic combinations of these 2-point blocks produce 3-point structures transforming homogeneously under superinversion:
\begin{align}
{\frak X}_{1}=X_{12+}^{-1}X_{23+}X_{31-}^{-1}\to -X_{1-}{\frak X}_{1}X_{1+},
\label{eq:3ptblocksboson3d}
\end{align}
and similarly for ${\frak X}_{2},{\frak X}_{3}$.

Using the 2-point blocks together with the $\theta$-coordinates, one can also construct purely Grassmannian 3-point structures \cite{Park99,AAN13}
\begin{align}
\Theta_{1\alpha}^{a}=(X_{21+}^{-1}\theta_{21}^a)_{\alpha}-(X_{31+}^{-1}\theta_{31}^a)_{\alpha},
\label{eq:3ptblocksgrass3d}
\end{align}
with $\Theta_{2},\Theta_{3}$ defined similarly. These structures are charged under $R$-symmetry and vanish in the non-supersymmetric limit $\theta\to 0$. Under superinversion they transform as \cite{AJ22}
\begin{align}
\Theta_{i\alpha}^{a}\to -(X_{i-})_{\alpha}^{\ \beta}\Theta_{i\beta}^{b}V_{ib}^{\ a},\qquad
\Theta_{i}^{a\alpha}\to (V^{T}_{i})^{a}_{\ b}\Theta_{i}^{b\beta}(X_{i+})_{\beta}^{\ \alpha}.
\end{align}
Here, $V_{j},\ j=1,2,3$ is a matrix acting on the $R$-symmetry indices \cite{Park99,BKS15,AJ22}
\begin{align}
V_{ja}^{\ b}=\delta_{a}^{\ b}+i\theta_{jb}X_{j+}^{-1}\theta^{a}_{j}.
\label{eq:Vmatrix3d}
\end{align}
Note that $V_{j}^{T}=V_{j}^{-1}$, giving us $V_{j}^{T}V_{j}=1$. This $V$-matrix is precisely the object that makes the $\Theta$-structures $R$-symmetry covariants rather than invariants: constructing true superconformal invariants from these building blocks requires contracting free $R$-symmetry indices using the invariant tensors of $SO(\cN)$, which for $\cN\geq 3$ includes the antisymmetric tensor $\epsilon^{a_{1}\hdots a_{\cN}}$. This is the origin of the $\epsilon$-invariants that are the central novel feature of the $\cN=3,4$ analysis carried out in section~\ref{sec:invariants}.

\section{Construction of superconformal invariants}
\label{sec:invariants}

For 3d CFTs, Giombi et al developed the auxiliary polarization spinor formalism to build conformally invariant structures that describe the form of spinning 3-point conformal correlators \cite{GPY11}. Their approach had been calibrated for superconformal theories for 3d $\cN=1$ SCFTs in \cite{AAN13} and for 3d $\cN=2$  SCFTs in \cite{AJ22}. In this section, we build superconformally invariant structures through a similar approach incorporating the non-abelian $R$-symmetry of $\cN=3,4$ SCFTs. 

We augment superspace with auxiliary commuting $SL(2,\mathbb{R})$ spinors $\lambda^{\alpha}$, one for each spacetime point:
\begin{align}
\hat z^{A}\equiv \{x^{\M}, \theta^{a\A},\lambda^{\B}\},\quad
\begin{tabular}{l}
$\M=0,1,2,$\\
$\A,\B=1,2,\quad a=1,2,\hdots,\cN$.
\end{tabular}
\end{align}
A spin-$s$ traceless symmetric superfield operator $\cO^{\A_{1}\hdots \A_{2s}}(x,\theta)$ is made index-free by contraction with $\lambda$'s:
\begin{align}
\cO_{s}(x,\theta,\LL)=\cO^{\A_{1}\hdots \A_{2s}}(x,\theta)\,\LL_{\A_{1}}\hdots \LL_{\A_{2s}}.
\end{align}
Since $\LL$'s are commuting and null ($\LL^{\A}\LL_{\A}=0$), the index-free operator $\cO_{s}$ remains traceless and symmetric, transforms as a Lorentz scalar, and carries scaling dimension $\Delta-s$. Superconformal invariants are then constructed by contracting $\lambda_{i}$'s and/or the Grassmannian structures $\Theta_{i}^{a}$ (defined in section~\ref{sec:covbuilding}) with the covariant 2- and 3-point building blocks, arranged to transform invariantly under super-Poincar\'e transformations and under superinversion up to a sign. Structures which pick up a minus sign under superinversion re classified as \textit{parity-odd}. This strategy was developed in \cite{GPY11} and extended to superconformal theories in \cite{AAN13,AJ22,AJ24}.

Structures built from $\lambda$'s alone are the supersymmetric extensions of the conformal invariants of \cite{GPY11}:
\begin{align}
\begin{aligned}
\text{(parity-even)}&\qquad P_{1}=\lambda_{2}X_{23-}^{-1}\lambda_{3},\quad
Q_{1}=\lambda_{1}X_{12-}^{-1}X_{23+}X_{31-}^{-1}\lambda_{1}\quad\text{and perm},\\
\text{(parity-odd)}&\qquad 
S_{1}=\frac{\lambda_{3}X_{31+}X_{12+}\lambda_{2}}{|\bar x_{12}||\bar x_{23}||\bar x_{31}|}\quad\text{and perm.}
\end{aligned}
\label{eq:bosonicinv}
\end{align}
These `bosonic' superconformal invariants were first constructed for 3d $\cN=1$ SCFTs in \cite{AAN13} and remain valid for all $\cN$ \cite{AJ22}.

\subsection{Fermionic covariants and $\cN=2$ invariants}
\label{sec:fermionic}

Structures built using $\Theta$'s carry free $R$-symmetry indices and are therefore superconformally \textit{covariant} rather than invariant; we refer to them as \textit{fermionic} structures since they vanish in the non-supersymmetric limit $\theta\to 0$. The $R$-symmetry covariants obtained\footnote{Note that we have defined the covariants here with $\xi_i$'s to get rid of the scaling factors after superinversion.} for $\cN=2$ in \cite{AJ22} continue to hold for $\cN=3,4$; we list them together with their superinversion transformations:
\begin{enumerate}[label=$\bullet$]
	\item $R_{i}^{a}=\lambda_{i}\Theta_{i}^{a}\to R_{i}^{b}(V_{i})_{b}^{\ a}$
	\item $\displaystyle\pi_{ij}^{a}=\frac{1}{\xi_{i}}\lambda_{i}X_{ij+}\Theta_{j}^{a}
	\to\pi_{ij}^{b} (V_{j})_{b}^{\ a}$
	\item $\sigma_{13}^{a}=\xi_{3}\dfrac{\lambda_{1}X_{12+}X_{23+}\Theta_{3}^{a}}
	{\sqrt{\bar x_{12}^{2}\bar x_{23}^{2}\bar x_{31}^{2}}}\to \sigma_{13}^{b}(V_{3})_{b}^{\ a}$
	\item $\omega^{a}_{i}={\xi_{i}}\lambda_{i}\mathfrak{X}_{i+}\Theta_{i}^{a}
	\to -\omega_{i}^{b}(V_{i})_{b}^{\ a}$	
	\item $\Psi_{i}^{ab}=\xi_{i}\displaystyle\Theta^{a}_{i\alpha}\Theta^{a\alpha}_{i}
	\to-(V^{T}_{i})^{a}_{\ c}\Psi_{i}^{cd}(V_{i})_{d}^{\ b}$
	\item $\Pi_{ij}^{ab}=\Theta_{i}^{a}X_{ij+}\Theta^{b}_{j}
	\to (V^{T}_{i})^{a}_{\ c}\Pi_{ij}^{cd}(V_{j})_{d}^{\ b}$
	\item $\Sigma_{13}^{ab}=\xi_{1}\xi_{3}\dfrac{\Theta^{a}_{1}X_{12+}X_{23+}\Theta^{b}_{3}}
	{\sqrt{\bar x_{12}^{2}\bar x_{23}^{2}\bar x_{31}^{2}}}
	\to - (V^{T}_{1})^{a}_{\ c}\Sigma^{cd}_{13}(V_{3})_{d}^{\ b}$
	\item $\Omega_{i}^{ab}=\xi_{i}^{2}\Theta_{i}^{a}\mathfrak{X}_{i+}\Theta_{i}^{b}
	\to(V^{T}_{i})^{a}_{\ c}\Omega_{i}^{cd}(V_{i})_{d}^{\ b}$
\end{enumerate}
Here, $\xi_1=\frac{\bar x_{12}\bar x_{31}}{\bar x_{23}}$, with other permutations defined similarly, is the normalization factor defined to have unit-normalized superinversion transformations. The free $R$-symmetry indices in these structures are contracted using the invariant tensors of $SO(\cN)$ --- namely $\delta^{ab}$ for all $\cN$, and $\epsilon^{a_{1}\cdots a_{\cN}}$ for $\cN\geq 3$, to produce superconformally invariant structures. For $\cN=1$, the $V$-matrix is trivial and all covariants are true invariants \cite{AAN13}. For $\cN=2$, contracting with $\delta^{ab}$ yields the fermionic invariants \cite{AJ22}:
\begin{align}
\begin{aligned}
\text{(parity-even)}&\qquad \bar R_{1}\equiv \pi_{21}^{a}\pi^{a}_{31}\ 
\text{and perm},\quad R'\equiv -2i \Omega_{1}^{aa},\\
\text{(parity-odd)}&\qquad T'\equiv \Psi^{aa}_1.
\end{aligned}
\end{align}
The complete set of $\cN=2$ invariants $\{P_i,Q_i,S_i,\bar R_i,R',T'\}$ remains valid for $\cN=3,4$. For $\cN\geq 3$, contracting with $\epsilon^{a_1\cdots a_\cN}$ produces additional invariants not present for $\cN\leq 2$, which we construct next.

\subsection{Novel $\epsilon$-invariants for $\cN=3$}
\label{sec:epsilonn3}

For $\cN=3$, the $R$-symmetry group is $SO(3)$, which admits a completely antisymmetric invariant tensor $\epsilon^{abc}$. Contracting three copies of the same $\Theta_{i}$ with $\epsilon^{abc}$:
\begin{align*}
U\sim\epsilon^{abc}\Theta_{i}^{a\A}\Theta_{i}^{b\B}\Theta_{i}^{c\gamma},
\end{align*}
and closing the spinor indices with $\lambda$'s or $\epsilon_{\alpha\beta}$ produces new superconformally invariant structures, which we call \textit{$\epsilon$-invariants}. The requirement that these transform under superinversion invariantly up to a sign forces all three $\Theta$'s to carry the same point index $i$. The free spinor indices can then be closed either with a single $\lambda$ and $\epsilon^{\alpha\beta}$, or with three $\lambda$'s, giving invariants at $\lambda$-orders $O(\LL_{i})$, $O(\LL_{i}^{3})$, $O(\LL_{i}^{2}\LL_{j})$, and 
$O(\LL_{1}\LL_{2}\LL_{3})$.

A systematic analysis, detailed in appendix~\ref{app:epsilonrelsn3}, reveals that at each $\lambda$-order there is precisely one independent parity-even and one independent parity-odd $\epsilon$-invariant, up to permutations. Expressed in terms of the covariants above, these are:\footnote{The $R$-symmetry indices are obvious, and hence suppressed to avoid clutter.}
\begin{align}
\begin{aligned}
O(\lambda_{1}):\qquad\qquad&
	\begin{aligned}
	\text{parity-even:\qquad}& \epsilon\,\Omega_{1}R_{1}\\
	\text{parity-odd:\qquad}& \epsilon\,\Omega_{1}\omega_{1}
	\end{aligned}\\
O(\LL_{1}^{3}):\qquad\qquad&
	\begin{aligned}
	\text{parity-even:\qquad}& \epsilon\,R_{1}R_{1}R_{1}\\
	\text{parity-odd:\qquad}& \epsilon\,R_{1}R_{1}\omega_{1}
	\end{aligned}\\
O(\LL_{1}^{2}\LL_{2}):\qquad\qquad&
	\begin{aligned}
	\text{parity-even:\qquad}& \epsilon\,R_{1}R_{1}\sigma_{21}\\
	\text{parity-odd:\qquad}& \epsilon\,R_{1}R_{1}\pi_{21}
	\end{aligned}\\
O(\LL_{1}\LL_{2}\LL_{3}):\qquad\qquad&
	\begin{aligned}
	\text{parity-even:\qquad}& \epsilon\,R_{1}\sigma_{21}\sigma_{31}\\
	\text{parity-odd:\qquad}& \epsilon\,\omega_{1}\pi_{21}\pi_{31}
	\end{aligned}
\end{aligned}
\end{align}
We denote the parity-even $\epsilon$-invariants as $U_{ijk}$ and the parity-odd ones as $\tilde U_{ijk}$, where $ijk$ encodes the $\lambda$-homogeneity $\LL_{1}^{i}\LL_{2}^{j}\LL_{3}^{k}$. For example, $U_{210}$ is the structure at $O(\LL_{1}^{2}\LL_{2})$.\footnote{The indices $i,j,k$ on $\epsilon$-invariants denote $\lambda$-homogeneity, not a point label as for $P_i,Q_i,S_i$.} Including all permutations, the $\cN=3$ theory has 13 parity-even and 13 parity-odd $\epsilon$-invariants:
\begin{enumerate}[label=$\bullet$,itemsep=0em]
\item $O(\lambda_{i}):$ $U_{100}$ and 2 perm, $\tilde U_{100}$ and 2 perm
\item $O(\lambda_{i}^{3}):$ $U_{300}$ and 2 perm, $\tilde U_{300}$ and 2 perm
\item $O(\lambda_{i}^{2}\lambda_{j}):$ $U_{210}$ and 5 perm, $\tilde U_{210}$ and 5 perm
\item $O(\lambda_{1}\lambda_{2}\lambda_{3})$: $U_{111},\ \tilde U_{111}$
\end{enumerate}

\subsection{Novel $\epsilon$-invariants for $\cN=4$}
\label{sec:epsilonn4}

For $\cN=4$, the $R$-symmetry is $SO(4)$ with antisymmetric tensor $\epsilon^{abcd}$. The construction proceeds analogously, now using four copies of the same $\Theta_{i}$:
\begin{align*}
U\sim \epsilon^{abcd} \Theta_{i}^{a\A}\Theta_{i}^{b\B}\Theta_{i}^{c\gamma}\Theta_{i}^{d\delta}. 
\end{align*}
Since four $\Theta$'s contribute four spinor indices, these must be closed with combinations of zero, two, or four $\lambda$'s, giving $\lambda$-orders $O(1)$, $O(\LL_{i}^{2})$, $O(\LL_{i}\LL_{j})$, $O(\LL_{i}^{4})$, $O(\LL_{i}^{3}\LL_{j})$, $O(\LL_{i}^{2}\LL_{j}^{2})$, $O(\LL_{i}^{2}\LL_{j}\LL_{k})$. Again, there is precisely one independent parity-even and one parity-odd $\epsilon$-invariant at each order:
\begin{align}
\begin{aligned}
O(\lambda_{1}^{2}):\qquad\qquad&
	\begin{aligned}
	\text{parity-even:\qquad}&\epsilon\,\Omega_{1}R_{1}R_{1},\\
	\text{parity-odd:\qquad}&\epsilon\,\Omega_{1}R_{1}\omega_{1},
	\end{aligned}\\
O(\lambda_{1}\lambda_{2}):\qquad\qquad&
	\begin{aligned}
	\text{parity-even:\qquad}&\epsilon\,\Omega_{1}R_{1}\sigma_{21},\\
	\text{parity-odd:\qquad}&\epsilon\,\Omega_{1}R_{1}\pi_{21},
	\end{aligned}\\
O(\lambda_{1}^{4}):\qquad\qquad&
	\begin{aligned}
	\text{parity-even:\qquad}&\epsilon\,R_{1}R_{1}R_{1}R_{1},\\
	\text{parity-odd:\qquad}&\epsilon\,R_{1}R_{1}R_{1}\omega_{1},
	\end{aligned}\\
O(\lambda_{1}^{3}\lambda_{2}):\qquad\qquad&
	\begin{aligned}
	\text{parity-even:\qquad}&\epsilon\,R_{1}R_{1}R_{1}\sigma_{21},\\
	\text{parity-odd:\qquad}&\epsilon\,R_{1}R_{1}R_{1}\pi_{21},
	\end{aligned}\\
O(\lambda_{1}^{2}\lambda_{2}^{2}):\qquad\qquad&
	\begin{aligned}
	\text{parity-even:\qquad}&\epsilon\,\omega_{1}\omega_{1}\pi_{21}\pi_{21},\\
	\text{parity-odd:\qquad}&
	\epsilon\,R_{1}R_{1}\pi_{21}\sigma_{21}-\epsilon\,R_{2}R_{2}\pi_{12}\sigma_{12},
	\end{aligned}\\
O(\lambda_{1}^{2}\lambda_{2}\lambda_{3}):\qquad\qquad&
	\begin{aligned}
	\text{parity-even:\qquad}&\epsilon\,R_{1}R_{1}\pi_{21}\pi_{31},\\
	\text{parity-odd:\qquad}&\epsilon\,R_{1}\omega_{1}\pi_{21}\pi_{31}.
	\end{aligned}
\end{aligned}
\end{align}
Additionally, for $\cN=4$ there is a unique $\lambda$-free parity-even $\epsilon$-invariant,
\begin{align}
U'\equiv\epsilon\,\Omega_{1}\Omega_{1},
\end{align}
whose permutations are all identical, in analogy with the $\lambda$-free invariants $R'$ and $T'$. Including all permutations, the $\cN=4$ theory has 22 parity-even and 21 parity-odd $\epsilon$-invariants:
\begin{enumerate}[label=$\bullet$,itemsep=0em]
	\item $O(\text{no $\LL$})$: $U'$
	\item $O(\LL_{i}^{2}):$ $U_{200}$ and 2 perm, $\tilde U_{200}$ and 2 perm
	\item $O(\LL_{i}\LL_{j}):$ $U_{110}$ and 2 perm, $\tilde U_{110}$ and 2 perm
	\item $O(\LL_{i}^{4}):$ $U_{400}$ and 2 perm, $\tilde U_{400}$ and 2 perm
	\item $O(\LL_{i}^{2}\LL_{j}^{2}):$ $U_{220}$ and 2 perm, $\tilde U_{220}$ and 2 perm
	\item $O(\LL_{i}^{3}\LL_{j}):$ $U_{310}$ and 5 perm, $\tilde U_{310}$ and 5 perm
	\item $O(\LL_{i}^{2}\LL_{j}\LL_{k}):$ $U_{211}$ and 2 perm, $\tilde U_{211}$ and 2 perm
\end{enumerate}

\subsection{Relations amongst invariants}
\label{sec:relations}

The complete set of superconformal invariants for $\cN=3,4$ SCFTs in 3d is:
\begin{align}
\renewcommand{\arraystretch}{1.2}
\begin{tabular}{|c|c|c|}
\hline
& parity-even & parity-odd\\\hline
bosonic & $P_{i}, Q_{i}$ & $S_{i}$\\\hline
fermionic & $\bar R_{i}, R'$ & $T'$\\\hline
$\epsilon$-invariants & $U_{ijk}$ & $\tilde U_{ijk}$\\\hline
\end{tabular}
\label{eq:scftinvariantsfinal}
\end{align}
with the $\epsilon$-invariants classified separately for each $\cN$. While exhaustive, these invariants are not independent and satisfy non-linear relations due to the nilpotency of fermionic structures and the algebraic properties of $\epsilon$-tensors. These relations have direct consequences for the structure of 3-point correlators and we summarize them here.

The simplest general relation is that a product of two parity-odd invariants is either zero or a sum of parity-even invariants:
\begin{align}
\left(\text{parity-odd}\right)\cdot\left(\text{parity-odd}\right)
=\sum\left(\text{parity-even}\right).
\label{eq:oddoddeven}
\end{align}
This holds for the bosonic invariants, the fermionic invariants, and the $\epsilon$-invariants alike. Examples valid for both $\cN=3,4$ include
\begin{align}
\begin{gathered}
S_2 S_3 + P_2 P_3 - Q_1 P_1 = 0, \qquad
S_1 T' + 2\bar R_1 = 0,
\end{gathered}
\end{align}
while for the $\cN=3$ $\epsilon$-invariants,
\begin{align}
\begin{gathered}
\tilde U_{300} S_1 + U_{210} P_2 - U_{201} P_3 = 0, \qquad
\tilde U_{100} \tilde U_{021} + \bar R_3 P_1 R' + 3\bar R_1 P_3 R'= 0,
\end{gathered}
\end{align}
and for $\cN=4$,
\begin{align}
\begin{gathered}
\tilde U_{200} S_1 + U_{110} P_2 - U_{101} P_3 = 0, \qquad
\tilde U_{110} \tilde U_{002} - 3P_3 Q_3 R'^2 + 6P_1 P_2 R'^2 = 0.
\end{gathered}
\end{align}
It follows from~\eqref{eq:oddoddeven} that each parity-odd invariant can appear at most linearly in any monomial describing a 3-point correlator.

\paragraph{Products of $\cN=2$ invariants}

Many of the non-linear relations among $\{P_i,Q_i,\bar R_i,R'\}$ derived for $\cN=2$ in \cite{AJ22} continue to hold for $\cN=3,4$. The lowest-order ones are:
\begin{gather}
\sum_{i=1}^{3}P_{i}^{2}Q_{i}-2P_{1}P_{2}P_{3}-Q_{1}Q_{2}Q_{3}
+\frac{i}{2}\sum_{i=1}^{3}\bar R_{i}P_{i}Q_{i}
+\frac{1}{4}P_{1}P_{2}P_{3}R'=0,\label{eq:n2rel1}\\
P_{1}\bar R_{2}+P_{2}\bar R_{1}+Q_{3}\bar R_{3}-\frac{i}{2}P_{1}P_{2}R'=0,
\quad(\text{and perm})\label{eq:n2rel2}
\end{gather}
and products of $\bar R_i$'s reduce as
\begin{gather}
2\bar R_{3}^{2}+\left(Q_{1}Q_{2}-P_{3}^{2}+iP_{3}\bar R_{3}\right)R'=0,\qquad
2\bar R_{1}\bar R_{2}+\left(P_{1}P_{2}-Q_{3}P_{3}\right)R'=0.
\end{gather}

However, not all $\cN=2$ relations survive. A notable example is
\begin{align}
(R_{1}^{a}R_{2}^{a})^{2}+\frac{1}{2}\left(Q_{1}Q_{2}-P_{3}^{2}\right)R'\ 
\begin{cases}
= 0, & \cN=2,\\
\neq 0, & \cN=3,4.
\end{cases}
\end{align}
This is a consequence of the enlarged superspace: since a fermionic structure can contain at most $\theta^{2\cN}$ in a 3d $\cN$-extended theory,\footnote{This is also why any fermionic structure multiplied with $R'\,[\Theta^{4}]$ vanishes for $\cN=2$ \cite{AJ22}, whereas for $\cN=3$ the products $\bar R_i R',\,T'R'\,[\Theta^{6}]$ survive, and for $\cN=4$ even $R'^2\,[\Theta^{8}]$ is non-vanishing.} the lowest-surviving power of $R_{i}^{a}R_{j}^{a}$ grows with $\cN$. The correct 
relations for $\cN=3,4$ are:
\begin{align}
\text{$\cN=3$:}\quad&
(R_{1}^{a}R_{2}^{a})^{3}+\tfrac{1}{2}\left(Q_{1}Q_{2}-P_{3}^{2}\right)\bar R_{3}R'=0,\\
\text{$\cN=4$:}\quad&
(R_{1}^{a}R_{2}^{a})^{4}-\left(\tfrac{1}{4}Q_{1}^{2}Q_{2}^{2}
-\tfrac{1}{2}Q_{1}Q_{2}P_{3}^{2}+\tfrac{1}{4}P_{3}^{4}\right)R'^{2}=0,
\end{align}
and analogous relations hold for $(\pi_{ij}^{a}\pi_{kl}^{a})$, $(\sigma_{ij}^{a}\sigma_{kl}^{a})$, $(\omega_{i}^{a}\omega_{j}^{a})$, and other bilinear covariants listed in \cite{AJ22}.

\paragraph{Products of $\epsilon$-invariants}

Since a product of two $\epsilon$-tensors decomposes into products of $\delta$'s, products of $\epsilon$-invariants reduce to combinations of $\cN=2$ invariants.

For $\cN=3$, all squares vanish,
\begin{align}
U_{ijk}^{2} = 0, \qquad \tilde U_{ijk}^{2} = 0,
\label{eq:N3esq}
\end{align}
and general products satisfy
\begin{align}
U_{ijk}\, U_{lmn} = \sum c_i\,\mathcal{B}_i\,\bar R_k R', \qquad
U_{ijk}\, \tilde U_{lmn} = \sum c_i\,\mathcal{B}_i\, R'T',
\label{eq:N3erelations}
\end{align}
where $\mathcal{B}_i$ are monomials in $\{P_i,Q_j\}$. For example,
\begin{align}
\begin{gathered}
U_{100}\, U_{010} - 4\bar R_3 R' = 0, \qquad
U_{100}\, U_{210} + Q_1\bar R_3 R' = 0,\\
U_{100}\, \tilde U_{010} - 2P_3 R'T' = 0.
\end{gathered}
\end{align}

For $\cN=4$, the squares vanish only for the highest-$\lambda$-order invariants,
\begin{align}
U_{ijk}^{2} = 0,\quad \tilde U_{ijk}^{2} = 0,\qquad (ijk\in\{400,310\}\ \text{and perms}),
\label{eq:N4esq}
\end{align}
while remaining products reduce to monomials in $\{P_i,Q_j\}$ multiplied by $R'^2$ 
(for even $\times$ even) or by $S_k R'^2$ (for even $\times$ odd):
\begin{align}
U_{ijk}\, U_{lmn} = \sum c_i\,\mathcal{B}_i\, R'^{2}, \qquad
U_{ijk}\, \tilde U_{lmn} = \sum c_i\,\mathcal{B}_i\, S_k R'^{2}.
\label{eq:N4erelations}
\end{align}
Examples include:
\begin{align}
\begin{gathered}
U'^{2} - 16R'^{2} = 0, \qquad
\tilde U_{020}^{2} - 3Q_2^2 R'^{2} = 0, \\
U_{220}^{2} + \left(-Q_1^2 Q_2^2 - P_3^4 + 2Q_1 Q_2 P_3^2\right) R'^{2} = 0, \\
\tilde U_{220}^{2} + \left(3Q_1^2 Q_2^2 - 3Q_1 Q_2 P_3^2\right) R'^{2} = 0,\\
U_{400}\, U_{020} + \left(6Q_1^2 Q_2 - 6Q_1 P_3^2\right) R'^{2} = 0, \qquad
\tilde U_{400}\, \tilde U_{200} + \tfrac{3}{2} Q_1^3 R'^{2} = 0,\\
\tilde U_{002}\, U_{200} - 6P_2 S_2 R'^{2} = 0, \qquad
\tilde U_{110}\, U_{200} + Q_1 S_3 R'^{2} = 0.
\end{gathered}
\end{align}
The complete list of such relations is given in appendix~\ref{app:epsilonrels}.

Together, \eqref{eq:N3esq}--\eqref{eq:N3erelations} and \eqref{eq:N4esq}--\eqref{eq:N4erelations} imply that $\epsilon$-invariants, like parity-odd invariants, can appear at most linearly in any monomial describing a 3-point correlator.

\subsection{Permutation symmetry}
\label{sec:permsym}

The structural form of a 3-point correlator is also constrained by imposing (anti)symmetry under exchange of a pair of operators, called permutation or point-switch (anti)symmetry. For a correlator $\langle \cO_i \cO_i \cO_j\rangle$ respecting symmetry under the exchange $(x_1,\theta_1,\lambda_1)\leftrightarrow (x_2,\theta_2,\lambda_2)$ (which we henceforth refer to as a $1\leftrightarrow2$ swap), the monomial structures describing it must also each remain invariant after the swap. The building blocks transform as
\begin{align}
X_{12+} \; \left\{
\begin{aligned}
&\xrightarrow{\;1\leftrightarrow2\;} X_{21+} \\
&\xrightarrow{\;2\leftrightarrow3\;} X_{13+} \\
&\xrightarrow{\;3\leftrightarrow1\;} X_{32+}
\end{aligned}
\right.,\qquad
\Theta_{1} \; \left\{
\begin{aligned}
&\xrightarrow{\;1\leftrightarrow2\;} -\Theta_{2}\\
&\xrightarrow{\;2\leftrightarrow3\;} -\Theta_{1} \\
&\xrightarrow{\;3\leftrightarrow1\;} -\Theta_{3}
\end{aligned}
\right.,\qquad
\lambda_{1} \; \left\{
\begin{aligned}
&\xrightarrow{\;1\leftrightarrow2\;} \lambda_{2}\\
&\xrightarrow{\;2\leftrightarrow3\;} \lambda_{1} \\
&\xrightarrow{\;3\leftrightarrow1\;} \lambda_{3}
\end{aligned}
\right.
\end{align}
and similarly for inverses and scalars. From these, the transformations of all superconformal invariants follow directly.

\paragraph{Bosonic and fermionic invariants}

The bosonic invariants $P_i, Q_i, S_i$ and the fermionic invariants $\bar R_i$ all transform in the same way,
\begin{align}
A_{1} \; \left\{
\begin{aligned}
&\xrightarrow{\;1\leftrightarrow2\;} -A_{2}\\
&\xrightarrow{\;2\leftrightarrow3\;} -A_{1} \\
&\xrightarrow{\;3\leftrightarrow1\;} -A_{3}
\end{aligned}
\right.,
\end{align}
while $R', T'$ are invariant under any swap.

\paragraph{$\epsilon$-invariants}

Under point swaps, an $\epsilon$-invariant can behave in one of three ways:
\begin{align}
A_{ijk} \; \left\{
\begin{aligned}
&\xrightarrow{\;1\leftrightarrow2\;}\, A_{jik}\\
&\xrightarrow{\;2\leftrightarrow3\;}\, A_{ikj} \\
&\xrightarrow{\;3\leftrightarrow1\;}\, A_{kji}
\end{aligned}\right.,\qquad
B_{ijk} \; \left\{
\begin{aligned}
&\xrightarrow{\;1\leftrightarrow2\;}\, -A_{jik}\\
&\xrightarrow{\;2\leftrightarrow3\;}\, -A_{ikj} \\
&\xrightarrow{\;3\leftrightarrow1\;}\, -A_{kji}
\end{aligned}\right.,\qquad
C_{ijk} \xrightarrow{\;\text{any swap}\;} C_{ijk}
\end{align}
The classification for all $\cN=3$ and $\cN=4$ $\epsilon$-invariants is as follows:
\begin{center}
\renewcommand{\arraystretch}{1.3}
\begin{tabular}{|c|c|c|}
\hline
 & $\cN=3$ & $\cN=4$ \\
\hline
$\ A_{ijk}\ $ & $U_{100}$ and perms & $U_{400}$ and perms,\ $U_{310}$ and perms,\\
              & $\tilde U_{300}$ and perms                & $U_{220}$ and perms,\ $U_{211}$ and perms,\ \\
\hline
$\ B_{ijk}$\  & $U_{300}$ and perms,\ $U_{210}$ and perms & $U_{200}$ and perms,\ $U_{110}$ and perms\\
              & $\tilde U_{100}$ and perms,\ $\tilde U_{210}$ and perms & $\tilde U_{400}$ and perms,\ $\tilde U_{310}$ and perms\\
              &                                           & $\tilde U_{220}$ and perms,\ $\tilde U_{211}$ and perms\\
              &                                           & $\tilde U_{200}$ and perms,\ $\tilde U_{110}$ and perms\\
\hline
$\ C_{ijk}\ $ & $U_{111}$,\ $\tilde U_{111}$              & $U'$ \\
\hline
\end{tabular}
\end{center}

Note that since the $\cN=3$ $\epsilon$-invariants contain three $\Theta$'s, they anticommute when permuted through each other, e.g. 
\begin{align}
U_{100}U_{010}\xrightarrow{\;1\leftrightarrow2\;} U_{010}U_{100} = -U_{100}U_{010},
\end{align}
consistent with the product relation $U_{100}U_{010}\propto \bar R_{3}R'$ and $\bar R_{3}R'\xrightarrow{\;1\leftrightarrow2\;}-\bar R_{3}R'$.

With these transformation rules in hand, we now turn to the construction of 3-point correlators in the next section.

\section{3-point SCFT correlators}
\label{sec:correlators}

The complete list of superconformal invariants in \eqref{eq:scftinvariantsfinal}, together with the relations and permutation symmetry constraints of sections~\ref{sec:relations} and~\ref{sec:permsym}, determines the structural form of all 3-point correlators in $\cN=3,4$ SCFTs. A 3-point correlator of index-free operators $\cO_{s_{i}}$ with spins $s_{i}$ and scaling dimensions $\Delta_{i}$ takes the form 
\begin{align}
\langle\cO_{s_{1}}\cO_{s_{2}}\cO_{s_{3}}\rangle
=\frac{1}{|\bar x_{12}|^{\tau_{12,3}}|\bar x_{23}|^{\tau_{23,1}}|\bar x_{31}|^{\tau_{31,2}}}
\left(\sum_{m}a_{m}{\cal T}^{\rm even}_{m}+\sum_{n}b_{n}{\cal T}^{\rm odd}_{n}\right),
\label{eq:finalcorrelator}
\end{align}
where $\tau_{ij,k}=\tau_{i}+\tau_{j}-\tau_{k}$ and $\tau_{i}=\Delta_{i}-s_{i}$. Each monomial ${\cal T}^{\rm even}_{m}$ (${\cal T}^{\rm odd}_{n}$) is parity-even (parity-odd) and homogeneous in $\lambda_{i}$ with degree $2s_{i}$. From the results of sections~\ref{sec:relations} and~\ref{sec:permsym}: each parity-odd monomial contains exactly one parity-odd invariant, and each $\epsilon$-invariant appears at most linearly. Furthermore, since $\epsilon$-invariants carry an odd number of $\lambda$'s for $\cN=3$ (and an even number for $\cN=4$), they contribute only to correlators with half-integer (integer) total spin for $\cN=3$ ($\cN=4$).

For conserved superfield operators $\cJ^{\A_{1}\hdots \A_{2s}}(x,\theta)$, the shortening condition
\begin{align}
D_{\A_1}\cJ^{\A_{1}\hdots \A_{2s}}(x,\theta)=0
\end{align}
imposes additional constraints on the correlator. In index-free notation this reads \cite{AAN13}
\begin{align}
\frac{\partial}{\partial \LL_{\alpha}}D_{\A}\cJ_{s}(x,\theta,\LL)=0,
\end{align}
and additionally fixes the conformal dimension to $\Delta=s+1$, i.e. $\tau=1$. Imposing these constraints on one or more operators in \eqref{eq:finalcorrelator} yields linear relations among $\{a_m, b_n\}$, reducing the correlator to a handful of undetermined OPE coefficients.

We present the results separately for $\cN=3$ and $\cN=4$. For each correlator, we first list the allowed structures for the non-conserved correlator, then state the result after imposing conservation on the relevant operators. We use `$\sim$' to denote the presence of the kinematic prefactor in \eqref{eq:finalcorrelator}.

\subsection{Correlators in $\cN=3$ SCFTs}
\label{sec:n3corr}

\paragraph{$\langle \cO_{0}\cO_{0}\cO_{0}\rangle$}

The correlator has no $\LL$'s, and the possible structures are
\begin{align*}
\text{parity-even:}\quad& 
1,\ R',
\\
\text{parity-odd:}\quad&
T', R'T'.
\end{align*}
Note that the correlator possesses permutation symmetry under $1\leftrightarrow2\leftrightarrow3$, which is preserved by all the listed structures (refer section~\ref{sec:permsym}). The 3-point correlator has the form
\begin{align}
\langle \cO_{0}\cO_{0}\cO_{0}\rangle \sim\  a_{1}+ a_{2} R'+b_{1}T'+b_{2}T'R'.
\end{align}
In the non-supersymmetric limit the fermionic structures vanish, leaving a single OPE coefficient $a_1$, as expected.

\paragraph{$\langle \cO_{\frac12}\cO_{0}\cO_{0}\rangle$} 

The correlator has homogeneity $\lambda_{1}$. It is symmetric under a $2\leftrightarrow 3$ swap. The possible 3-point structures only involve the $\epsilon$-invariants:
\begin{align*}
\text{parity-even:}\quad& 
U_{100}.
\end{align*}
The parity-odd structure $\tilde U_{100}$ is also possible but it does not satisfy the $2\leftrightarrow3$ permutation symmetry (refer section~\ref{sec:permsym}), and is consequently removed. The allowed 3-point correlator takes the form
\begin{align}
\langle \cO_{\frac12}\cO_{0}\cO_{0}\rangle \sim\  a U_{100}.
\end{align}
Hence, the correlator is fixed up to an overall constant. Conservation on $\cO_{\frac12}$ gives the only constraint
\begin{align}
\Delta_{2}=\Delta_{3}=\Delta.
\end{align}
Thus, the final form of the correlator becomes
\begin{align}
\langle
\cJ_{\frac12}\cO_{0}\cO_{0}\rangle=\frac{a}{|\bar x_{12}||\bar x_{23}|^{2\Delta-1}|x_{31}|}U_{100}.
\end{align}

\paragraph{$\langle \cO_{\frac32}\cO_{0}\cO_{0}\rangle$}

The homogeneity is $\LL_{1}^{3}$, and only parity-odd structures are symmetric under $2\leftrightarrow3$ swap:
\begin{align*}
\text{parity-odd:}\quad&
\tilde U_{300},\ \tilde U_{100}Q_{1}.
\end{align*}
The parity-even structures $U_{300},\ U_{100}Q_{1}$ are removed since they are antisymmetric under $2\leftrightarrow3$ swap. The correlator looks like
\begin{align}
\langle \cO_{\frac32}\cO_{0}\cO_{0}\rangle\sim \ b_{1}\tilde U_{300}+b_{2}\tilde U_{100}Q_{1}.
\end{align}
Employing conservation on $\cO_{\frac32}$ gives
\begin{align}
b_{1}=b_{2}=0,
\end{align}
i.e. the conserved correlator $\langle \cJ_{\frac32}\cO_{0}\cO_{0}\rangle$ vanishes.

\paragraph{$\langle \cO_{\frac52}\cO_{0}\cO_{0}\rangle$}

Only parity-even structures are allowed due to permutation symmetry:
\begin{align*}
\text{parity-even:}\quad& U_{300}Q_{1},\  U_{100}Q_{1}^{2}.
\end{align*}
The correlator takes the form
\begin{align}
\langle \cO_{\frac32}\cO_{0}\cO_{0}\rangle\sim \ a_{1} U_{300}Q_{1}+a_{2} U_{100}Q_{1}^{2}.
\end{align}
Conservation on $\cO_{\frac52}$ gives the relation
\begin{align}\textstyle
a_{2}=-\frac34 a_{1},\quad
\Delta_{2}=\Delta_{3},
\end{align}
i.e. the correlator is fixed up to an overall coefficient.

\paragraph{$\langle \cO_{n+\frac12}\cO_{0}\cO_{0}\rangle$}

For $n\geq1$ and odd, i.e. when the spin of the first operator is $s=\frac32, \frac72,\frac{11}2, \hdots$, only the parity-odd structures preserve the $2\leftrightarrow 3$ permutation symmetry of the correlator. There are only two allowed structures, obtained by multiplying $Q_{1}^{n-1}$ with the structures in $\langle \cO_{\frac32}\cO_{0}\cO_{0}\rangle$. When the operator $\cO_{n+\frac12}$ is conserved, the correlator vanishes.

For $n\geq2$ and even, i.e. for spins $s=\frac52, \frac92,\frac{13}2 \hdots$, the allowed permutation-symmetric structures are parity-even and are obtainable by multiplying $Q_{1}^{n-2}$ with the structures in $\langle \cO_{\frac52}\cO_{0}\cO_{0}\rangle$.

\paragraph{$\langle \cO_{1}\cO_{0}\cO_{0}\rangle$}

There are no structures which satisfy the permutation symmetry under a $2\leftrightarrow3$ swap, hence the correlator vanishes.

\paragraph{$\langle \cO_{2}\cO_{0}\cO_{0}\rangle$}

The allowed 3-point structures, symmetric under $2\leftrightarrow3$ swap are
\begin{align*}
\text{parity-even:}\quad& 
Q_{1}^{2},\ Q_{1}^{2}R',
\\
\text{parity-odd:}\quad&
Q_{1}^{2}T',\  Q_{1}^{2}R'T'.
\end{align*}
The correlator has the form
\begin{align}
\langle \cO_{2}\cO_{0}\cO_{0}\rangle\sim
Q_{1}^{2}\left(a_{1}+a_{2}R'+b_{1}T'+b_{2}T'R'\right).
\end{align}
When $\cO_{2}$ is conserved, we get the constraints
\begin{align}\textstyle
a_{2}=\frac{5}{16} a_{1},\quad
b_{2}=b_{1}=0,\quad
\Delta_{2}=\Delta_{3}.
\end{align}
Thus, $\langle \cJ_{2}\cO_{0}\cO_{0}\rangle$ is fixed up to a single parity-even coefficient, and does not contain a parity-odd contribution.

\paragraph{$\langle \cO_{s}\cO_{0}\cO_{0}\rangle$}

For $s>2$ and odd, there are no structures which preserve the permutation symmetry of the correlator.

For $s>2$ and even, the allowed structures are all parity-even and can be obtained by multiplying $Q_{1}^{s-2}$ with the structures for $\langle \cO_{2}\cO_{0}\cO_{0}\rangle$. Furthermore, imposing conservation on $\cO_{s}$ gives 
\begin{align}\textstyle
a_{2}=\frac{2s+1}{16} a_{1},\quad
b_{2}=b_{1}=0,\quad
\Delta_{2}=\Delta_{3},
\end{align}
i.e. the correlator $\langle \cJ_{2}\cO_{0}\cO_{0}\rangle$ is fixed up to an overall coefficient. Note that the correlator matches exactly with the $\cN=2$ result \cite{AJ22}.

\paragraph{$\langle \cO_{\frac12}\cO_{\frac12}\cO_{0}\rangle$}

The homogeneity of the correlator is $\LL_{1}\LL_{2}$, and the allowed structures  are anti-symmetric under $1\leftrightarrow2$ swap:
\begin{align*}
\text{parity-even:}\quad& 
P_{3},\ P_{3}R',\ \bar R_{3},\ \bar R_{3}R'
\\
\text{parity-odd:}\quad&
P_{3}T',\ P_{3}R'T',\ S_{3},\ S_{3}R'.
\end{align*}
The structure $\bar R_{3}T'$ is a possible parity-odd structure, but it is linearly dependent on the others. The 3-point correlator looks like
\begin{align}
\begin{aligned}
\langle \cO_{\frac12}\cO_{\frac12}\cO_{0}\rangle&\sim
P_{3}(a_{1}+a_{2} R')+ \bar R_{3}(a_{3}+a_{4} R')\\
&+P_{3} T'(b_{1}+b_{2} R')+S_{3}(b_{3}+b_{4} R').
\end{aligned}
\end{align}
When the first $\cO_{\frac12}$ is conserved, we obtain the relations (define $\delta_{23}=\Delta_2-\Delta_3$)
\begin{align}
\begin{gathered}\textstyle
a_{4}=\frac{-(\delta_{23}^{2}-9)(\delta_{23}-1)}{96}a_{1},\ 
a_{3}=\frac{i(\delta_{23}-1)}{2}a_{1},\ 
a_{2}=\frac{-(\delta_{23}-1)(\delta_{23}+3)}{16}a_{1},\ 
\\\textstyle
b_{4}=\frac{-i(\delta_{23}-2)}{4}b_{1},\ 
b_{3}=\frac{4i}{(\delta_{23}+2)}b_{1},\ 
b_{2}=\frac{-(\delta_{23}-2)}{48}b_{1}.
\end{gathered}
\end{align}
Similarly, when the second $\cO_{\frac12}$ is conserved, we get identical relations with $1\leftrightarrow2$. Thus, the correlator is determined up to one parity-even and one parity-odd structures.               

\paragraph{$\langle \cO_{\frac32}\cO_{\frac12}\cO_{0}\rangle$}

The correlator with the allowed structures is
\begin{align}
\begin{aligned}
\langle \cO_{\frac32}\cO_{\frac12}\cO_{0}\rangle&\sim
 Q_{1}P_{3}(a_{1}+a_{2}R')+Q_{1}\bar R_{3}(a_{3}+a_{4}R')\\
&+Q_{1}S_{3}(b_{1}+b_{2}R')+Q_{1}P_{3}T'(b_{3}+b_{4}R').
\end{aligned}
\end{align}
Upon imposing conservation on $\cO_{\frac12}$, one gets ($\delta_{13}=\Delta_1-\Delta_3$)
\begin{align}
\begin{gathered}\textstyle
a_{2}=\frac{-(2\delta_{13}-5)(2\delta_{13}+7)}{64}a_{1},\quad
a_{3}=\frac{i(2\delta_{13}-5)}{4}a_{1},\quad
a_{4}=\frac{-i(2\delta_{13}-9)(2\delta_{13}-5)(2\delta_{13}+7)}{768}a_{1},
\\\textstyle
b_{2}=\frac{-(2\delta_{13}-7)(2\delta_{13}+9)}{192}b_{1},\quad
b_{3}=\frac{8i}{2\delta_{13}+5}b_{1},\quad
b_{4}=\frac{-i(2\delta_{13}-7)}{8}b_{1}.
\end{gathered}
\end{align}
Upon further imposing conservation on $\cO_{\frac32}$, the parity-odd contributions drop out:
\begin{align}
\begin{gathered}\textstyle
a_{2}=\frac{-\Delta_{3}(\Delta_{3}-6)}{16}a_{1},\quad
a_{3}=\frac{-i\Delta_{3}}{2}a_{1},\quad
a_{4}=\frac{i\Delta_{3}(\Delta_{3}-6)(\Delta_{3}+2)}{96}a_{1},
\\\textstyle
b_{2}=
b_{3}=
b_{4}=b_{1}=0.
\end{gathered}
\end{align}
Thus, the correlator $\langle \cJ_{\frac32}\cJ_{\frac12}\cO_{0}\rangle$ is fixed up to a single parity-even structure. 

\paragraph{$\langle \cO_{1}\cO_{\frac12}\cO_{0}\rangle$}

The correlator has the structures
\begin{align}
\begin{aligned}
\langle \cO_{1}\cO_{\frac12}\cO_{0}\rangle& \sim 
a_{1}U_{010}Q_{1}+ a_{2}U_{100}P_{3}+a_{3} U_{210}+ b_{1}U_{010}Q_{1}+b_{2} \tilde U_{100}P_{3}+b_{3} \tilde U_{210}.
\end{aligned}
\end{align}
When $\cO_1,\cO_\frac12$ are conserved, we get
\begin{align}
a_2=2a_1,\ a_3=-2a_1,\quad b_3=b_2=b_1=0.
\end{align}
i.e. the correlator is fixed with one parity-even structure and there are no surviving parity-odd structures.

\paragraph{$\langle \cO_{s}\cO_{\frac12}\cO_{0}\rangle$}
For $s>1$, the structures are obtainable by multiplying $Q_{1}^{s-1}$ with the 3-point structures of $\langle \cO_{1}\cO_{\frac12}\cO_{0}\rangle$.

\paragraph{$\langle \cO_{1}\cO_{1}\cO_{0}\rangle$}

The non-conserved correlator consists of the structures
\begin{align}
\begin{aligned}
\langle \cO_{1}\cO_{1}\cO_{0}\rangle&\sim
Q_{1}Q_{2}(a_{1}+a_{2}R')+P_{3}^{2}(a_{3}+a_{4}R')+P_{3}\bar R_{3}(a_{5}+a_{6}R')\\
&+P_{3}S_{3}(b_{1}+b_{2}R')+Q_{1}Q_{2}T'(b_{3}+b_{4}R')+P_{3}^{2}T'(b_{5}+b_{6}R').
\end{aligned}
\end{align}
Upon imposing conservation on just the first operator $\cO_1$, we get
\begin{align}
\begin{gathered}\textstyle
a_{2}=\frac{-(\delta_{23}-4)(\delta_{23}+2)}{16}a_{1},\quad
a_{3}=\frac{4-2(\delta_{23}-2)}{\delta_{23}-2}a_{1},\quad
a_{4}=\frac{\delta_{23}(\delta_{23}+2)}{8}a_{1},
\\\textstyle
a_{5}=-i(\delta_{23}+1)a_{1},\quad
a_{6}=\frac{i(\delta_{23}-4)(\delta_{23}+1)(\delta_{23}+2)}{48}a_{1},
\\\textstyle
b_{2}=\frac{-(\delta_{23}-3)(\delta_{23}+1)}{16}b_{1},\quad
b_{3}=\frac{i(\delta_{23}-3)}{8}b_{1},\quad
b_{4}=\frac{-i(\delta_{23}^{2}-9)(\delta_{23}-5)}{384}b_{1},
\\\textstyle
b_{5}=\frac{-i\delta_{23}}{4}b_{1},\quad
b_{6}=\frac{i(\delta_{23}-3)\delta_{23}(\delta_{23}+3)}{192}b_{1}.
\end{gathered}
\end{align}
We get similar relations when conservation is instead imposed on the second $\cO_1$, with the obvious replacement $\Delta_2\to\Delta_1$.

Thus, the relations for the correlator $\langle \cJ_{1}\cJ_{1}\cO_{0}\rangle$ are
\begin{align}
\begin{gathered}\textstyle
a_{2}=\frac{-(\Delta_{3}-4)(\Delta_{3}+2)}{16}a_{1},\quad
a_{3}=\frac{4-2\Delta_3}{\Delta_{3}}a_{1},\quad
a_{4}=\frac{(\Delta_{3}-4)(\Delta_{3}-2)}{8}a_{1},
\\\textstyle
a_{5}=i(\Delta_{3}-3)\,a_{1},\quad
a_{6}=\frac{-i(\Delta_{3}-4)(\Delta_{3}-3)(\Delta_{3}+2)}{48}a_{1},
\\\textstyle
b_{2}=\frac{-(\Delta_{3}-3)(\Delta_{3}+1)}{16}b_{1},\quad
b_{3}=\frac{-i(\Delta_{3}+1)}{8}b_{1},\quad
b_{4}=\frac{i(\Delta_{3}-5)(\Delta_{3}+1)(\Delta_{3}+3)}{384}b_{1},
\\\textstyle
b_{5}=\frac{i(\Delta_{3}-2)}{4}b_{1},\quad
b_{6}=\frac{-i(\Delta_{3}-5)(\Delta_{3}-2)(\Delta_{3}+1)}{192}b_{1}.
\end{gathered}
\end{align}
The correlator is fixed up to a single parity-even and a single parity-odd structure.
\paragraph{$\langle \cO_{s}\cO_{1}\cO_{0}\rangle$}

For $s\geq2$, the allowed structures are $Q_{1}^{s-1}$ times the structures for $\langle \cO_{1}\cO_{1}\cO_{0}\rangle$.

\paragraph{$\langle \cO_{2}\cO_{2}\cO_{0}\rangle$}

The correlator has the structures
\begin{align}
\begin{aligned}
\langle \cO_{2}\cO_{2}\cO_{0}\rangle &\sim
Q_{1}^{2}Q_{2}^{2}(a_{11}+a_{12}R')
+Q_{1}Q_{2}P_{3}^{2}(a_{12}+a_{22}R')
+P_{3}^{4}(a_{31}+a_{32}R')\\
&+Q_{1}Q_{2}P_{3}\bar R_{3}(a_{41}+a_{42}R')
+P_{3}^{3}\bar R_{3}(a_{51}+a_{52}R')\\
&+Q_{1}^{2}Q_{2}^{2}T'(b_{11}+b_{12}R')
+Q_{1}Q_{2}P_{3}^{2}T'(b_{21}+b_{22}R')
+P_{3}^{4}T'(b_{31}+b_{32}R')\\
&+Q_{1}Q_{2}P_{3}S_{3}(b_{41}+b_{42}R')
+P_{3}^{3}S_{3}(b_{51}+b_{52}R').
\end{aligned}
\end{align}
When the second operator $\cO_2$ is conserved, we get ($\delta_3\equiv\Delta_1-\Delta_3$):
\begin{align}
\begin{gathered}\textstyle
a_{12}=\frac{-(\delta_{13}-7)(\delta_{13}+3)}{16}a_{11},\ 
a_{21}=\frac{8(\delta_{13}-5)+48}{\delta_{13}-5}a_{11},\  a_{22}=\frac{\delta_{13}^{2}+\delta_{13}+1}{2}a_{11},
\\\textstyle
a_{31}=8\left(1+\frac{3}{\delta_{13}-3}-\frac{11}{\delta_{13}-5}\right)a_{11},\ 
a_{32}=\frac{-(\delta_{13}^{2}-5)(\delta_{13}+1)}{2(\delta_{13}-5)}a_{11},\ 
a_{41}=-2i(\delta_{13}+2)\,a_{11},
\\\textstyle
a_{42}=\frac{i(\delta_{13}-7)(\delta_{13}+2)(\delta_{13}+3)}{24}a_{11},\ 
a_{51}=\frac{4i(\delta_{13}-1)(\delta_{13}+2)}{\delta_{13}-5}a_{11},\ 
a_{52}=\frac{-i(\delta_{13}^{2}-3)(\delta_{13}+2)}{12}a_{11},
\\\textstyle
b_{12}=\frac{-(\delta_{13}-8)(\delta_{13}+4)}{48}b_{11},\ 
b_{21}=\frac{-8(\delta_{13}-6)+52}{\delta_{13}-6}b_{11},\ 
b_{22}=\frac{2\delta_{13}^2+7\delta_{13}-7\delta_{13}+2}{12}b_{11},
\\\textstyle
b_{31}=8\left(1+\frac{5}{\delta_{13}-4}-\frac{14}{\delta_{13}-6}\right)b_{11},\ 
b_{32}=\frac{-(\delta_{13}^2-6)(\delta_{13}+1)}{6(\delta_{13}-6)}b_{11},\ b_{41}=\frac{16i}{\delta_{13}-6}b_{11},
\\\textstyle
b_{42}=i(\delta_{13}+2)\,b_{11},\ 
b_{51}=\frac{32i\,\delta_{13}}{(\delta_{13}-6)(\delta_{13}-4)}b_{11},\ 
b_{52}=\frac{-2i(\delta_{13}^2-2)}{\delta_{13}-6}b_{11}.
\end{gathered}
\end{align}
Thus, the correlator with both the spin-2 operators conserved has the relations:
\begin{align}
\begin{gathered}\textstyle
a_{12}=\frac{-(\Delta_3-6)(\Delta_3+4)}{16}a_{11},\ 
a_{21}=\frac{-8(\Delta_3-4)}{\Delta_3+2}a_{11},\ 
a_{22}=\frac{\Delta_3^2-9\Delta_3+19}{2}a_{11},
\\\textstyle
a_{31}=8\left(\frac{3}{\Delta_3}+\frac{\Delta_3-9}{\Delta_3+2}\right)a_{11},\ 
a_{32}=\frac{-(\Delta_3-4)(\Delta_3^2-6\Delta_3+4)}{2(\Delta_3+2)}a_{11},\ a_{41}=2i(\Delta_3-5)a_{11},
\\\textstyle
a_{42}=\frac{-i(\Delta_3-6)(\Delta_3-5)(\Delta_3+4)}{24}a_{11},\ 
a_{51}=\frac{-4i(\Delta_3-5)(\Delta_3-2)}{\Delta_3+2}a_{11},\ 
a_{52}=\frac{i(\Delta_3-5)(\Delta_3^2-6\Delta_3+6)}{12}a_{11},
\\\textstyle
b_{12}=\frac{-(\Delta_3-7)(\Delta_3+5)}{48}b_{11},\ 
b_{21}=\frac{4(7-2\Delta_3)}{\Delta_3+3}b_{11},\ 
b_{22}=\frac{2\Delta_3^2-19\Delta_3+41}{12}b_{11},
\\\textstyle
b_{31}=\frac{8(\Delta_3-4)(\Delta_3-1)}{(\Delta_3+1)(\Delta_3+3)}b_{11},\ 
b_{32}=\frac{-(\Delta_3-4)(\Delta_3^2-6\Delta_3+3)}{6(\Delta_3+3)}b_{11},\ b_{41}=\frac{16i}{\Delta_3+3}b_{11},
\\\textstyle
b_{42}=-i(\Delta_3-5)b_{11},\ 
b_{51}=\frac{-32i(\Delta_3-3)}{(\Delta_3+1)(\Delta_3+3)}b_{11},\ 
b_{52}=\frac{2i(\Delta_3^2-6\Delta_3+7)}{\Delta_3+3}b_{11}.
\end{gathered}
\end{align}
The final conserved correlator $\langle \cJ_{2}\cJ_{2}\cO_{0}\rangle$ is fixed up to a single parity-even and a single-parity-odd structure.

\paragraph{$\langle \cO_{s}\cO_{2}\cO_{0}\rangle$}

For $s>2$ the 3-point structures are obtainable by multiplying $Q_{1}^{s-2}$ with structures in $\langle \cO_{2}\cO_{2}\cO_{0}\rangle$.

\paragraph{$\langle \cO_{\frac12}\cO_{\frac12}\cO_{\frac12}\rangle$}

The correlator is fully antisymmetric under $1\leftrightarrow2\leftrightarrow3$ and 
only parity-even structures are allowed. The correlator takes the form
\begin{align}
\langle \cO_{\frac12}\cO_{\frac12}\cO_{\frac12}\rangle\sim\ 
a_{1}(U_{100}P_{1}+U_{010}P_{2}+U_{001}P_{3})+a_{2}\,U_{111}.
\end{align}
Upon conservation on any of the operators, we get the relation
\begin{align}
a_2=-2a_1,\quad \Delta_1=\Delta_2=\Delta_3=\tfrac32.
\end{align}

\paragraph{$\langle \cO_{\frac32}\cO_{\frac12}\cO_{\frac12}\rangle$}

The correlator has a $2\leftrightarrow3$ permutation antisymmetry and takes the form
\begin{align}
\langle \cO_{\frac32}\cO_{\frac12}\cO_{\frac12}\rangle\sim\ 
a_{1}U_{300}P_{1}
+a_{2}(U_{210}P_{2}+U_{201}P_{3})
+b_{1}(\tilde U_{210}P_{2}+\tilde U_{201}P_{3}).
\end{align}

\paragraph{$\langle \cO_{\frac72}\cO_{\frac12}\cO_{\frac12}\rangle$}

The allowed structures are $Q_{1}^{2}$ times those of $\langle \cO_{\frac32}\cO_{\frac12}\cO_{\frac12}\rangle$, and the correlator takes 
the form
\begin{align}
\langle \cO_{\frac72}\cO_{\frac12}\cO_{\frac12}\rangle\sim\ 
a_{1}Q_{1}^{2}U_{300}P_{1}
+a_{2}Q_{1}^{2}(U_{210}P_{2}+U_{201}P_{3})
+b_{1}Q_{1}^{2}(\tilde U_{210}P_{2}+\tilde U_{201}P_{3}).
\end{align}

For correlators of the form $\langle \cO_{\frac s2}\cO_{\frac12}\cO_{\frac12}\rangle$, conservation on any of the operators results in the vanishing of the conserved correlator.

\paragraph{$\langle \cO_{1}\cO_{\frac12}\cO_{\frac12}\rangle$}

The correlator has a $2\leftrightarrow3$ permutation antisymmetry, which severely 
restricts the allowed structures. The correlator looks like
\begin{align}
\langle \cO_{1}\cO_{\frac12}\cO_{\frac12}\rangle\sim\ 
(P_{2}\bar R_{3}-P_{3}\bar R_{2})(a_{1}+a_{2}R')
+(P_{2}S_{3}-P_{3}S_{2})(b_{1}+b_{2}R').
\end{align}
The correlator does not survive conservation on any of the operators, hence the conserved correlator vanishes.

\paragraph{$\langle \cO_{2}\cO_{\frac12}\cO_{\frac12}\rangle$}

The correlator contains the structures
\begin{align}
\begin{aligned}
\langle \cO_{2}\cO_{\frac12}\cO_{\frac12}\rangle&\sim
Q_{1}^{2}P_{1}(a_{11}+a_{12}R')
+Q_{1}^{2}\bar R_{1}(a_{21}+a_{22}R')
+Q_{1}P_{2}P_{3}(a_{31}+a_{32}R')\\
&+Q_{1}^{2}S_{1}(b_{11}+b_{12}R')
+Q_{1}^{2}P_{1}T'(b_{21}+b_{22}R')
+Q_{1}P_{2}P_{3}T'(b_{31}+b_{32}R').
\end{aligned}
\end{align}
Upon imposing conservation on the first $\cO_{\frac12}$, we get ($\delta_{13}\equiv\Delta_1-\Delta_3$):
\begin{align}
\begin{gathered}\textstyle
a_{12}=\frac{-(2\delta_{13}-9)(2\delta_{13}+7)}{64}a_{11},\ 
a_{21}=\frac{-i(2\delta_{13}+7)}{4}a_{11},\ 
a_{22}=\frac{i(2\delta_{13}-9)(2\delta_{13}+7)(2\delta_{13}+11)}{768}a_{11},\\\textstyle
a_{31}=\frac{8}{2\delta_{13}-5}a_{11},\ 
a_{32}=\frac{-(2\delta_{13}+7)}{8}a_{11},\\\textstyle
b_{12}=\frac{-(2\delta_{13}-7)(2\delta_{13}+9)}{64}b_{11},\ 
b_{21}=\frac{i(2\delta_{13}-7)}{8}b_{11},\\\textstyle
b_{22}=\frac{-i(2\delta_{13}-11)(2\delta_{13}-7)(2\delta_{13}+9)}{1536}b_{11},\ 
b_{31}=i\,b_{11},\ 
b_{32}=\frac{-i(2\delta_{13}-7)(2\delta_{13}+9)}{192}b_{11}.
\end{gathered}
\end{align}

\paragraph{$\langle \cO_{s}\cO_{\frac12}\cO_{\frac12}\rangle$}

For odd $s>2$, the 3-point structures are obtained by multiplying the structures of $\langle \cO_{1}\cO_{\frac12}\cO_{\frac12}\rangle$ with $Q_{1}^{s-1}$. 

Similarly, for even $s>2$, the structures are $Q_{1}^{s-2}$ times the structures of $\langle \cO_{2}\cO_{\frac12}\cO_{\frac12}\rangle$.

\paragraph{$\langle \cO_{1}\cO_{1}\cO_{1}\rangle$}

Due to the permutation symmetry of the correlator under any of the $1\leftrightarrow2\leftrightarrow3$ swaps, there are no structures possible and the 3-point correlator vanishes.

\paragraph{$\langle \cO_{2}\cO_{1}\cO_{1}\rangle$}

The correlator has a $2\leftrightarrow3$ permutation symmetry and takes the form
\begin{align}
\begin{aligned}
\langle \cO_{2}\cO_{1}\cO_{1}\rangle&\sim
Q_{1}^{2}Q_{2}Q_{3}(a_{11}+a_{12}R')
+Q_{1}^{2}P_{1}^{2}(a_{21}+a_{22}R')
+Q_{1}^{2}P_{1}\bar R_{1}(a_{31}+a_{32}R')\\
&+Q_{1}P_{1}P_{2}P_{3}(a_{41}+a_{42}R')
+Q_{1}P_{2}P_{3}\bar R_{1}(a_{51}+a_{52}R')
+P_{2}^{2}P_{3}^{2}(a_{61}+a_{62}R')\\
&+Q_{1}^{2}P_{1}S_{1}(b_{11}+b_{12}R')
+Q_{1}P_{1}(P_{2}S_{2}+P_{3}S_{3})(b_{21}+b_{22}R')\\
&+Q_{1}P_{2}P_{3}S_{1}(b_{31}+b_{32}R')
+P_{2}P_{3}(P_{2}S_{2}+P_{3}S_{3})(b_{41}+b_{42}R')\\
&+Q_{1}^{2}Q_{2}Q_{3}T'(b_{51}+b_{52}R')
+Q_{1}^{2}P_{1}^{2}T'(b_{61}+b_{62}R').
\end{aligned}
\end{align}
Imposing conservation on the second operator, one finds ($\delta_{13}\equiv\Delta_1-\Delta_3$):
\begin{align}
\begin{gathered}\textstyle
a_{12}=\frac{-(\delta_{13}-4)(\delta_{13}+6)}{16}a_{11},\ 
a_{21}=\frac{-2\delta_{13}}{\delta_{13}+4}a_{11},\ 
a_{22}=\frac{(\delta_{13}-4)\delta_{13}}{8}a_{11},\ a_{31}=i(\delta_{13}-3)a_{11},
\\\textstyle
a_{32}=\frac{-i(\delta_{13}-4)(\delta_{13}-3)(\delta_{13}+6)}{48}a_{11},\ a_{41}=\frac{-8}{\delta_{13}+4}a_{11},\ a_{42}=\frac{\delta_{13}-4}{2}a_{11},\ a_{51}=4ia_{11},
\\\textstyle
a_{52}=\frac{-i(\delta_{13}-4)(\delta_{13}+6)}{12}a_{11},\ 
a_{61}=\frac{-8}{(\delta_{13}-2)(\delta_{13}+4)}a_{11},\ 
a_{62}=\frac{1}{2}a_{11},\\\textstyle
b_{12}=\frac{-(5\delta_{13}-19)(\delta_{13}+5)}{144}b_{11},\ 
b_{21}=\frac{2}{3}b_{11},\ 
b_{22}=\frac{-(\delta_{13}^{2}-25)}{72}b_{11},\ b_{31}=\frac{2}{\delta_{13}-3}b_{11},
\\\textstyle
b_{32}=\frac{-7(\delta_{13}+5)}{72}b_{11},\ 
b_{41}=\frac{-2}{3(\delta_{13}-3)}b_{11},\ 
b_{42}=\frac{-(\delta_{13}+5)}{72}b_{11},\ b_{51}=\frac{-i(\delta_{13}+5)}{24}b_{11},
\\\textstyle
b_{52}=\frac{i(\delta_{13}-5)(\delta_{13}+5)(\delta_{13}+7)}{1152}b_{11},\ 
b_{61}=\frac{i\delta_{13}}{12}b_{11},\ 
b_{62}=\frac{-i(\delta_{13}-5)\delta_{13}(\delta_{13}+5)}{576}b_{11}.
\end{gathered}
\end{align}
Alternatively, conservation on the third operator gives the same relations. Further imposing conservation on $\cO_2$ only fixes its conformal dimension. Thus, the conserved correlator is fixed with one parity-even and one parity-odd structures.









\subsection{Correlators in $\cN=4$ SCFTs}
\label{sec:n4corr}

The final list of superconformal invariants for $\cN=4$ is presented in~\eqref{eq:scftinvariantsfinal}. Note that the examples in $\cN=3$ greatly outnumber those in $\cN=4$ due to the presence of half-integer operators.

\paragraph{$\langle \cO_{0}\cO_{0}\cO_{0}\rangle$}

The allowed structures are:
\begin{align*}
\text{parity-even:}\quad& 
1,\ 
R',\ 
R'^2,\ 
U',
\\
\text{parity-odd:}\quad&
T',\ 
T'R'.
\end{align*}
i.e. the correlator looks like
\begin{align}
\langle \cO_{0}\cO_{0}\cO_{0}\rangle\sim\ 
a_{1}+a_{2}R'+a_{3}R'^{2}+a_{4}U'+b_{1}T'+b_{2}T'R'.
\end{align}

\paragraph{$\langle \cO_{1}\cO_{0}\cO_{0}\rangle$}

There are no structures which satisfy the permutation symmetry under a $2\leftrightarrow3$ swap, hence the correlator vanishes.

\paragraph{$\langle \cO_{2}\cO_{0}\cO_{0}\rangle$}

The correlator is symmetric under $2\leftrightarrow3$ swap and contains the structures
\begin{align}
\begin{aligned}
\langle \cO_{2}\cO_{0}\cO_{0}\rangle\sim\ &
Q_{1}^{2}(a_{1}+a_{2}R'+a_{3}R'^{2}+a_{4}U')
+a_{5}Q_{1}U_{200}+a_{6}U_{400}
\\
&+Q_{1}^{2}T'(b_{1}+b_{2}R').
\end{aligned}
\end{align}
When $\cO_{2}$ is conserved, we get 
\begin{align}\textstyle
a_{2}=\frac{5}{16}a_{1},\  a_{3}=\frac{35}{512}a_{1},\  a_{5}=-8a_{4},\  a_{6}=\frac{8}{3}a_{4}, \quad b_{1}=b_{2}=0,\quad \Delta_{2}=\Delta_{3}.
\end{align}
The parity-even part of the conserved correlator is fixed to \textit{two} independent structures, parameterized by $a_1$ and $a_4$ respectively. This is in contrast to the $\cN=1,2,3$ cases \cite{AAN13,AJ22}, where conservation leaves a single parity-even structure. The second structure, parameterized by $a_4$, arises entirely from the $\epsilon$-invariants. The parity-odd sector vanishes identically upon conservation, consistent with the triangle inequality of the spins.

\paragraph{$\langle \cO_{s}\cO_{0}\cO_{0}\rangle$}

For $s>2$ and odd, there are no structures which preserve the permutation symmetry. 

For $s>2$ and even, the allowed structures are parity-even and are obtained by multiplying $Q_{1}^{s-2}$ with the structures for $\langle \cO_{2}\cO_{0}\cO_{0}\rangle$.

\paragraph{$\langle \cO_{1}\cO_{1}\cO_{0}\rangle$}
The correlator contains the following structures
\begin{align}
\begin{aligned}
\langle \cO_{1}\cO_{1}\cO_{0}\rangle &\sim
Q_{1}Q_{2}(a_{11}+a_{12}R'+a_{13}R'^{2}+a_{14}U')\\
&+P_{3}^{2}(a_{21}+a_{22}R'+a_{23}R'^{2}+a_{24}U')\\
&+P_{3}\bar R_{3}(a_{31}+a_{32}R')
+a_{4}P_{3}U_{110}+a_{5}U_{220}\\
&+Q_{1}Q_{2}T'(b_{11}+b_{12}R')
+P_{3}^{2}T'(b_{21}+b_{22}R')\\
&+P_{3}S_{3}(b_{31}+b_{32}R'+b_{33}R'^{2}+b_{34}U')+b_{4}P_{3}\tilde U_{110}+b_{5}S_{3}U_{110}.
\end{aligned}
\end{align}
When the first $\cO_1$ is conserved, we get the relations ($\delta_{23}=\Delta_2-\Delta_3$)
\begin{align}
\begin{gathered}\textstyle
a_{12}=\frac{-(\delta_{23}-4)(\delta_{23}+2)}{16}a_{11},\ 
a_{13}=\frac{(\delta_{23}-6)(\delta_{23}-4)(\delta_{23}+2)(\delta_{23}+4)}{1536}a_{11},
\\\textstyle
a_{21}=\frac{4-2(\delta_{23}-2)}{\delta_{23}-2}a_{11},\ 
a_{22}=\frac{\delta_{23}(\delta_{23}+2)}{8}a_{11},\ 
a_{23}=\frac{-(\delta_{23}-4)\delta_{23}(\delta_{23}+2)(\delta_{23}+4)}{768}a_{11},
\\\textstyle
a_{31}=-i(\delta_{23}+1)a_{11},\ 
a_{32}=\frac{i(\delta_{23}-4)(\delta_{23}+1)(\delta_{23}+2)}{48}a_{11},
\\\textstyle
b_{12}=\frac{-(\delta_{23}-5)(\delta_{23}+3)}{48}b_{11},\ 
b_{21}=\frac{6-2(\delta_{23}-3)}{\delta_{23}-3}b_{11},\ 
b_{22}=\frac{\delta_{23}(\delta_{23}+3)}{24}b_{11},
\\\textstyle
b_{31}=\frac{8i}{\delta_{23}-3}b_{11},\ 
b_{32}=\frac{i(\delta_{23}+1)}{2}b_{11},\ 
b_{33}=\frac{-i(\delta_{23}-5)(\delta_{23}+1)(\delta_{23}+3)}{192}b_{11},\\
a_{14}=a_{24}=a_4=a_5=b_{34}=b_{4}=b_{5}=0,
\end{gathered}
\end{align}
A similar set of relations is obtained if alternatively the second operator is conserved. These results suggest that the parity-even $\epsilon$-terms do not survive the conservation constraint and the conserved correlator is fixed in terms of one parity-even and one parity-odd structure. Interestingly, the fully conserved correlator displays non-trivial behavior: when both the spin-1 operators are conserved, and the scalar operator has canonical dimensions ($\Delta_3=0$), we get the relations:
\begin{align}
\begin{gathered}\textstyle
a_{12}=\frac{9}{16}a_{11},\ 
a_{13}=\frac{75}{512}a_{11},\ 
a_{21}=2\,a_{11},\ 
a_{22}=\frac{3}{8}a_{11},\ 
a_{23}=\frac{15}{256}a_{11},\ 
a_{24}=-2\,a_{14},
\\\textstyle
a_{31}=-2i\,a_{11},\ 
a_{32}=\frac{-3i}{8}a_{11},\ 
a_{4}=-8\,a_{14},\ 
a_{5}=\frac{8}{5}a_{14},\\
\textstyle
b_{12}=\frac{1}{3}b_{11},\ 
b_{21}=b_{11},\ 
b_{22}=\frac{1}{6}b_{11},\\\textstyle
b_{31}=4i\,b_{11},\ 
b_{32}=i\,b_{11},\ 
b_{33}=\frac{i}{6}b_{11},\ 
b_{34}=b_{4}=b_{5}=0.
\end{gathered}
\end{align}
That is, the fully conserved correlator is fixed up to \textit{two} parity-even terms and a single parity-odd term. This trend is observed in many other $\cN=4$ examples, as we will note below.

\paragraph{$\langle \cO_{2}\cO_{1}\cO_{0}\rangle$}

The allowed structures are $Q_{1}$ times the structures for $\langle \cO_{1}\cO_{1}\cO_{0}\rangle$. Additionally, there are three more parity-even terms allowed: $a_{6} Q_{1}Q_2 U_{200}$, $a_7 P_3^2 U_{200}$, $a_8 Q_1^2 U_{020}$. When $\cO_2$ and $\cO_1$ are conserved, and the scalar has canonical dimensions, we get relations for the parity-even part:
\begin{align}
\begin{gathered}\textstyle
a_{12}=\frac{15}{16}a_{11},\ 
a_{13}=\frac{175}{512}a_{11},\ 
a_{21}=4\,a_{11},
\\\textstyle
a_{22}=\frac{5}{4}a_{11},\ 
a_{23}=\frac{35}{128}a_{11},\ 
a_{24}=\frac{-12}{25}a_{14},
\\\textstyle
a_{31}=-4i\,a_{11},\ 
a_{32}=\frac{-5i}{4}a_{11},\ 
a_{4}=\frac{24}{25}a_{14},\ 
a_{5}=\frac{16}{25}a_{14},\ 
a_{6}=\frac{28}{25}a_{14},
\\\textstyle
a_{7}=\frac{-8}{25}a_{14},\ 
a_{8}=\frac{-16}{25}a_{14},\ b_{ij}=0,
\end{gathered}
\end{align}
and the correlator is fixed by two parity-even structures with no parity-odd contribution.

\paragraph{$\langle \cO_{s}\cO_{1}\cO_{0}\rangle$}

For $s\geq2$, the 3-point structures are obtainable by multiplying $Q_{1}^{s-1}$ with the allowed structures for $\langle \cO_{1}\cO_{1}\cO_{0}\rangle$. An additional structure $a_{9} Q_1^{s-3}P_3^2 U_{400}$ is also allowed.  Imposing total conservation, we get for $s=3$:
\begin{align}
\begin{gathered}\textstyle
a_{12}=\frac{21}{16}a_{11},\ 
a_{13}=\frac{315}{512}a_{11},\ 
a_{21}=6\,a_{11},
\\\textstyle
a_{22}=\frac{21}{8}a_{11},\ 
a_{23}=\frac{189}{256}a_{11},\ 
a_{24}=\frac{-30}{79}a_{14},
\\\textstyle
a_{31}=-6i\,a_{11},\ 
a_{32}=\frac{-21i}{8}a_{11},\ 
a_{4}=\frac{60}{79}a_{14},\ 
a_{5}=\frac{80}{79}a_{14},\ 
a_{6}=\frac{84}{79}a_{14},
\\\textstyle
a_{7}=\frac{-40}{79}a_{14},\ 
a_{8}=\frac{-56}{79}a_{14},\ 
a_{9}=\frac{-16}{79}a_{14}.
\end{gathered}
\end{align}
For $s=4$:
\begin{align}
\begin{gathered}\textstyle
a_{12}=\frac{27}{16}a_{11},\ 
a_{13}=\frac{495}{512}a_{11},\ 
a_{21}=8\,a_{11},
\\\textstyle
a_{22}=\frac{9}{2}a_{11},\ 
a_{23}=\frac{99}{64}a_{11},\ 
a_{24}=\frac{-40}{127}a_{14},
\\\textstyle
a_{31}=-8i\,a_{11},\ 
a_{32}=\frac{-9i}{2}a_{11},\ 
a_{41}=\frac{80}{127}a_{14},\ 
a_{42}=\frac{160}{127}a_{14},\ 
a_{51}=\frac{132}{127}a_{14},
\\\textstyle
a_{61}=\frac{-80}{127}a_{14},\ 
a_{71}=\frac{-96}{127}a_{14},\ 
a_{81}=\frac{-64}{127}a_{14}.
\end{gathered}
\end{align}
The parity-odd part in these correlators (with $s>2$) does not survive any conservation constraint.

\paragraph{$\langle \cO_{2}\cO_{2}\cO_{0}\rangle$}

The correlator is
\begin{align}
\begin{aligned}
\langle \cO_{2}&\cO_{2}\cO_{0}\rangle\sim
Q_{1}^{2}Q_{2}^{2}(a_{11}+a_{12}R'+a_{13}R'^{2}+a_{14}U')+Q_{1}Q_{2}P_{3}^{2}(a_{21}+a_{22}R'+a_{23}R'^{2}+a_{24}U')\\
&+P_{3}^{4}(a_{31}+a_{32}R'+a_{33}R'^{2}+a_{34}U')+Q_{1}Q_{2}P_{3}\bar R_{3}(a_{41}+a_{42}R')\\
&+P_{3}^{3}\bar R_{3}(a_{51}+a_{52}R')+a_{61}Q_{1}Q_{2}P_{3}U_{110}+a_{62}P_{3}^{3}U_{110}+a_{71}Q_{1}Q_{2}U_{220}+a_{72}P_{3}^{2}U_{220}\\
&+Q_{1}^{2}Q_{2}^{2}T'(b_{11}+b_{12}R')
+Q_{1}Q_{2}P_{3}^{2}T'(b_{21}+b_{22}R')
+P_{3}^{4}T'(b_{31}+b_{32}R')\\
&+Q_{1}Q_{2}P_{3}S_{3}(b_{41}+b_{42}R'+b_{43}R'^{2})+P_{3}^{3}S_{3}(b_{51}+b_{52}R'+b_{53}R'^{2}).
\end{aligned}
\end{align}
When one of the spin-2 operators (say the second $\cO_2$) is conserved, we get
\begin{align}
\begin{gathered}\textstyle
a_{12}=\frac{-(\delta_{13}-7)(\delta_{13}+3)}{16}a_{11},\quad
a_{13}=\frac{(\delta_{13}-9)(\delta_{13}-7)(\delta_{13}+3)(\delta_{13}+5)}{1536}a_{11},\quad
a_{14}=0,
\\\textstyle
a_{21}=8\left(\frac{12-\delta_{13}}{\delta_{13}-5}\right)a_{11},\quad
a_{22}=\frac{\delta_{13}^{2}+3\delta_{13}+1}{2}a_{11},
\\\textstyle
a_{23}=\frac{-(\delta_{13}-7)(\delta_{13}+3)(\delta_{13}^{2}+4\delta_{13}+1)}{192}a_{11},\quad
a_{24}=0,
\\\textstyle
a_{31}=8\left(1+\frac{3}{\delta_{13}-3}-\frac{11}{\delta_{13}-5}\right)a_{11},\quad
a_{32}=\frac{-(\delta_{13}^{2}-5)(\delta_{13}+1)}{2(\delta_{13}-5)}a_{11},
\\\textstyle
a_{33}=\frac{(\delta_{13}^{2}-7)(\delta_{13}+1)(\delta_{13}+3)}{192}a_{11},\quad
a_{34}=0,
\\\textstyle
a_{41}=-2i(\delta_{13}+2)\,a_{11},\quad
a_{42}=\frac{i(\delta_{13}-7)(\delta_{13}+2)(\delta_{13}+3)}{24}a_{11},
\\\textstyle
a_{51}=\frac{4i(\delta_{13}-1)(\delta_{13}+2)}{\delta_{13}-5}a_{11},\quad
a_{52}=\frac{-i(\delta_{13}^{2}-3)(\delta_{13}+2)}{12}a_{11},\quad
a_{61}=a_{62}=a_{71}=a_{72}=0,
\\\textstyle
b_{12}=\frac{-(\delta_{13}-8)(\delta_{13}+4)}{48}b_{11},\quad
b_{21}=\frac{-8(\delta_{13}-6)+52}{\delta_{13}-6}b_{11},\quad
b_{22}=\frac{2\delta_{13}^{2}+7\delta_{13}+2}{12}b_{11},
\\\textstyle
b_{31}=8\left(1+\frac{5}{\delta_{13}-4}-\frac{14}{\delta_{13}-6}\right)b_{11},\quad
b_{32}=\frac{-(\delta_{13}^{2}-6)(\delta_{13}+1)}{6(\delta_{13}-6)}b_{11},
\\\textstyle
b_{41}=\frac{16i}{\delta_{13}-6}b_{11},\quad
b_{42}=i(\delta_{13}+2)\,b_{11},\quad
b_{43}=\frac{-i(\delta_{13}-8)(\delta_{13}+2)(\delta_{13}+4)}{96}b_{11},
\\\textstyle
b_{51}=\frac{32i\,\delta_{13}}{(\delta_{13}-6)(\delta_{13}-4)}b_{11},\quad
b_{52}=\frac{-2i(\delta_{13}^{2}-2)}{\delta_{13}-6}b_{11},\quad
b_{53}=\frac{i(\delta_{13}-2)(\delta_{13}+2)^{2}}{48}b_{11}.
\end{gathered}
\end{align}
When both the spinning operators are conserved and the scalar superfield operator has canonical dimensions $\Delta_3=1$, the parity-even part looks like:
\begin{align}
\begin{gathered}\textstyle
a_{12}=\frac{25}{16}a_{11},\ 
a_{13}=\frac{1225}{1536}a_{11},\ 
a_{21}=8\,a_{11},\ 
a_{22}=\frac{11}{2}a_{11},\ 
a_{23}=\frac{325}{192}a_{11},\ 
a_{24}=0,
\\\textstyle
a_{31}=\frac{8}{3}a_{11},\ 
a_{32}=\frac{-1}{2}a_{11},\ 
a_{33}=\frac{-15}{64}a_{11},\ 
a_{34}=-16\,a_{14},
\\\textstyle
a_{41}=-8i\,a_{11},\ 
a_{42}=\frac{-25i}{6}a_{11},\ 
a_{51}=\frac{-16i}{3}a_{11},\ 
a_{52}=\frac{-i}{3}a_{11},
\\\textstyle
a_{71}=-16\,a_{14},\ 
a_{72}=-32\,a_{14},\ 
a_{91}=\frac{16}{7}a_{14},\ 
a_{92}=\frac{96}{7}a_{14}.
\end{gathered}
\end{align}
Thus, for the fully conserved correlator, we obtain two independent parity-even structures. 

\paragraph{$\langle \cO_{3}\cO_{2}\cO_{0}\rangle$}

The 3-point structures are obtainable by multiplying $Q_{1}$ with the allowed structures for $\langle \cO_{2}\cO_{2}\cO_{0}\rangle$.

\paragraph{$\langle \cO_{1}\cO_{1}\cO_{1}\rangle$}

Due to the permutation symmetry of the correlator under any of the $1\leftrightarrow2\leftrightarrow3$ swaps and the antisymmetry of $\epsilon$-invariants under permutations, there are no allowed structures and the 3-point correlator vanishes.

\paragraph{$\langle \cO_{2}\cO_{1}\cO_{1}\rangle$}

The correlator has a $2\leftrightarrow3$ permutation symmetry and has the form
\begin{align}
\begin{aligned}
\langle \cO_{2}\cO_{1}\cO_{1}&\rangle\sim
Q_{1}^{2}Q_{2}Q_{3}(a_{11}+a_{12}R'+a_{13}R'^{2})
+Q_{1}^{2}P_{1}^{2}(a_{21}+a_{22}R'+a_{23}R'^{2})\\
&+Q_{1}P_{1}P_{2}P_{3}(a_{31}+a_{32}R'+a_{33}R'^{2})
+P_{2}^{2}P_{3}^{2}(a_{41}+a_{42}R'+a_{43}R'^{2})\\
&+a_{52}Q_{1}^{2}P_{1}\bar R_{1}R'
+Q_{1}P_{1}(P_{2}\bar R_{3}-P_{3}\bar R_{2})(a_{61}+a_{62}R')\\
&+P_{2}P_{3}(P_{2}\bar R_{3}-P_{3}\bar R_{2})(a_{71}+a_{72}R')\\
&+a_{901}Q_{1}^{2}Q_{2}Q_{3}U_{200}+a_{902}Q_{1}P_{1}^{2}U_{200}+a_{903}Q_{1}P_{1}P_{2}P_{3}U_{200}\\
&+a_{904}Q_{2}Q_{3}U_{400}+a_{905}P_{1}^{2}U_{400}\\
&+a_{906}(Q_{1}^{2}Q_{2}U_{002}+Q_{1}^{2}Q_{3}U_{020})
+a_{907}(Q_{1}P_{3}^{2}U_{002}+Q_{1}P_{2}^{2}U_{020})\\
&+a_{908}Q_{1}^{2}P_{1}U_{011}
+a_{909}(Q_{1}P_{1}P_{2}U_{110}+Q_{1}P_{1}P_{3}U_{101})\\
&+a_{910}(P_{2}^{2}P_{3}U_{110}+P_{3}^{2}P_{2}U_{101})
+a_{911}Q_{1}^{2}U_{022}\\
&+a_{912}Q_{1}P_{1}U_{211}+a_{913}P_{2}P_{3}U_{211}\\
&+Q_{1}^{2}Q_{2}Q_{3}T'(b_{12}+b_{13}R')
+Q_{1}^{2}P_{1}^{2}T'(b_{22}+b_{23}R')\\
&+Q_{1}P_{1}P_{2}P_{3}T'(b_{32}+b_{33}R')
+P_{2}^{2}P_{3}^{2}T'(b_{42}+b_{43}R')\\
&+Q_{1}^{2}P_{1}S_{1}(b_{51}+b_{52}R'+b_{53}R'^{2})
+Q_{1}P_{2}P_{3}S_{1}(b_{61}+b_{62}R'+b_{63}R'^{2}).
\end{aligned}
\end{align}
Upon imposing conservation on any one of the spin-1 operator (say the first), the correlator is fixed in terms of a single parity-even and a single parity-odd structure. On the other hand -- as seen in previous examples -- when all three operators are conserved, the correlator is fixed in terms of two independent parity-even and one parity-odd term (the parity-odd parts involving $\epsilon$ vanishes). We get the relations for the fully conserved correlator:
\begin{align}
\begin{gathered}\textstyle
a_{12}=\frac{21}{16}a_{11},\ 
a_{13}=\frac{315}{512}a_{11},\ 
a_{21}=\frac{-2}{5}a_{11},\ 
a_{22}=\frac{-3}{8}a_{11},\ 
a_{23}=\frac{-35}{256}a_{11},\\\textstyle
a_{31}=\frac{-8}{5}a_{11},\ 
a_{32}=\frac{-5}{2}a_{11},\ 
a_{33}=\frac{-35}{64}a_{11},\ 
a_{41}=\frac{8}{5}a_{11},\ 
a_{42}=\frac{5}{2}a_{11},\ 
a_{43}=\frac{63}{64}a_{11},\\\textstyle
a_{52}=\frac{-7i}{8}a_{11},\ 
a_{61}=2i\,a_{11},\ 
a_{62}=0,\ 
a_{71}=-4i\,a_{11},\ 
a_{72}=\frac{-7i}{4}a_{11}\\\textstyle
a_{902}=\frac{-27}{73}a_{901},\ 
a_{903}=\frac{182}{73}a_{901},\ 
a_{904}=\frac{18}{73}a_{901},\ 
a_{905}=\frac{-32}{73}a_{901},\ 
a_{906}=\frac{-13}{73}a_{901},\\\textstyle
a_{907}=\frac{23}{73}a_{901},\ 
a_{908}=\frac{40}{73}a_{901},\ 
a_{909}=\frac{66}{73}a_{901},\ 
a_{910}=\frac{14}{73}a_{901},\\\textstyle
a_{911}=\frac{-14}{73}a_{901},\ 
a_{912}=\frac{60}{73}a_{901},\ 
a_{913}=\frac{136}{73}a_{901}\\\textstyle
b_{13}=\frac{2}{3}b_{12},\ 
b_{22}=\frac{-1}{3}b_{12},\ 
b_{23}=\frac{-1}{6}b_{12},\ 
b_{32}=\frac{-4}{3}b_{12},\ 
b_{33}=\frac{-2}{3}b_{12},\\\textstyle
b_{42}=\frac{2}{3}b_{12},\ 
b_{43}=\frac{1}{6}b_{12},\ 
b_{51}=\frac{4i}{3}b_{12},\ 
b_{52}=i\,b_{12},\ 
b_{53}=\frac{i}{3}b_{12},\\\textstyle
b_{61}=\frac{-8i}{3}b_{12},\ 
b_{62}=-2i\,b_{12},\ 
b_{63}=\frac{-2i}{3}b_{12}.
\end{gathered}
\end{align}
The only undetermined coefficients are $a_{11}, a_{901}$ (parity-even) and $b_{12}$ (parity-odd).

\section{Discussion}
\label{sec:discussion}
In this work the constraints of superconformal invariance and current conservation on 3-point correlators of spinning superfield operators in $\cN=3$ and $\cN=4$ SCFTs in 3d were studied. Taking recourse to polarization spinor techniques for encoding spin and utilizing superspace methods we gave an exhaustive enumeration of superconformal invariants in $\cN=3$ and $\cN=4$ theories in 3d. This included the novel $\epsilon$-invariants that exist only for theories with non-abelian R-symmetry.

We then enumerated the structure of various 3-point correlators built out of these invariants and subsequently studied the further constraints arising out of current conservation. Applying conservation constraints to general spinning correlators, we find that for $\cN=3$, conserved 3-point functions of higher-spin currents are fixed by one parity-even and one parity-odd structure, in agreement with \cite{BKS15} for the supercurrent. 

For $\cN=4$, two parity-even structures and one parity-odd structure survive conservation. The second parity-even structure is a direct consequence of the $\epsilon$-invariants available for $SO(4)$ and has no analogue for $\cN\leq 3$. Physically, this second structure is associated with the breaking of mirror symmetry: the $R$-symmetry group $SO(4)\cong SU(2)_L\times SU(2)_R$ admits a $\mathbb{Z}_2$ mirror map exchanging $SU(2)_L\leftrightarrow SU(2)_R$, and the $\epsilon$-invariant structure in superspace is the lift of this $L$-$R$ asymmetry to the level of 3-point invariants. The first parity-even structure is present in all $\cN=4$ SCFTs, while the second is non-zero only in theories not invariant under the mirror map \cite{BKS15b}. This is consistent with the free hypermultiplet, which is mirror-symmetric and hence saturates only the first structure.

Together with the results of \cite{AAN13,AJ22} for $\cN=1,2$, this completes the systematic enumeration of spinning 3-point correlators in 3d $\cN$-extended SCFTs. As is well known, superspace methods are not expected to be of much utility for $\cN>4$ SCFTs in 3d.

Several directions for future work suggest themselves:

\begin{itemize}

\item {\it Generating functions.} Giombi, Prakash and Yin \cite{GPY11} presented generating functions for spinning 3-point correlators in the free boson and free fermion theories in 3d, which produce all conserved current correlators in those theories. The analogous generating functions for 3d SCFTs are not known for any $\cN$. Straightforward supersymmetric generalizations do not appear to work, and the structure of the $\epsilon$-invariants suggests that the answer may be qualitatively different for $\cN\geq3$.

\item {\it Superconformal bootstrap.} The 3-point functions of operators with arbitrary spin obtained here are a necessary ingredient for constructing superconformal blocks in $\cN=3$ and $\cN=4$ theories in 3d. These blocks would enable the numerical and analytical bootstrap program for 4-point functions in these theories, extending existing results for $\cN=1,2$ \cite{BS21,BS21b,BS23a,BS23b}.

\item {\it Maldacena-Zhiboedov type theorems for $\cN$-extended SCFTs.} In 3d CFTs, the existence of a single exactly conserved higher-spin operator implies the theory is free \cite{MZ1}, while weakly broken higher-spin symmetry tightly constrains correlators \cite{MZ2}. The supersymmetric analogues of these theorems are not known. Since the GPY formalism was instrumental in establishing \cite{MZ1,MZ2}, and comparable results in the same formalism are now available for 3d $\cN=1,2,3,4$ SCFTs, it is feasible that the supersymmetric generalization of the Maldacena-Zhiboedov theorems can now be explicated using similar techniques. 

\item {\it Extension to 4d $\cN>1$ SCFTs.} The program of enumerating spinning 3-point correlators via superspace and polarization spinor methods was carried out for 4d $\cN=1$ in \cite{AJ24}. Extending this to $\cN\ge2$ in 4d is a natural next step. The existence of the antisymmetric invariant tensor for the $R$-symmetry group $SU(\cN)$  suggests that $\epsilon$-invariants analogous to those constructed here will again play a central role. The $\cN=4$ case in 4d is of particular interest given the central role of $\mathcal{N}=4$ SYM in the AdS/CFT correspondence. The constraints of maximal supersymmetry are expected to fix correlators even more rigidly.

\end{itemize}

\acknowledgments

We acknowledge the use of the publicly available Mathematica package \textit{grassmann} by Matthew Headrick, which facilitated computations involving grassmanian variables. A.J. has been supported by an appointment to the JRG program at APCTP through the Science and Technology Promotion fund and Lottery fund of the Korean government.

\bigskip

\appendix

\section{Notation and conventions}
\label{app:conventions}

We use the mostly-plus metric $\eta_{\mu\nu}=\mathrm{diag}(-1,+1,+1)$. Spinors transform under the double cover of the 3d Lorentz group, $SL(2,\mathbb{R})$, and can be taken to be Majorana satisfying the reality condition $\psi_\alpha=\psi_\alpha^*$. Following \cite{Park99}, we work in the real basis for the gamma matrices, which throughout the paper are also written as $\sigma^\mu\equiv\gamma^\mu$; the Pauli matrices $\sigma^1,\sigma^2,\sigma^3$ appearing in the explicit representations below are distinguished by their Roman superscripts:
\begin{equation}
(\gamma_\mu)_\alpha^{\ \beta}\equiv(i\sigma^2,\,\sigma^1,\,\sigma^3)
=\left(\begin{pmatrix}0&1\\-1&0\end{pmatrix},\,
\begin{pmatrix}0&1\\1&0\end{pmatrix},\,
\begin{pmatrix}1&0\\0&-1\end{pmatrix}\right).
\end{equation}
With both spinor indices down or both up, the gamma matrices are symmetric:
\begin{equation}
(\gamma_\mu)_{\alpha\beta}=(\mathds{1},\,\sigma^3,\,-\sigma^1),\qquad
(\gamma_\mu)^{\alpha\beta}=(\mathds{1},\,-\sigma^3,\,\sigma^1).
\end{equation}
The antisymmetric epsilon symbol satisfies $\epsilon^{12}=-1=\epsilon_{21}$ and raises/lowers spinor indices as
\begin{equation}
\psi^\alpha=\epsilon^{\alpha\beta}\psi_\beta,\qquad\psi_\alpha=\epsilon_{\alpha\beta}\psi^\beta,
\end{equation}
with the shorthand $\psi\chi\equiv\psi^\alpha\chi_\alpha$. Multi-index objects are contracted following the NW-to-SE rule, e.g.\ $(AB)_\alpha{}^\gamma=A_\alpha{}^\beta B_\beta{}^\gamma$ and $\lambda X\mu\equiv\lambda^\alpha X_\alpha{}^\beta\mu_\beta$. The gamma matrices satisfy
\begin{equation}
(\gamma_\mu\gamma_\nu)_\alpha^{\ \beta}=\eta_{\mu\nu}\delta_\alpha^\beta+\epsilon_{\mu\nu\rho}(\gamma^\rho)_\alpha^{\ \beta},
\end{equation}
with the Levi-Civita convention $\epsilon_{012}=1$. 

The $R$-symmetry indices $a,b,\ldots=1,\ldots,\cN$ for $\cN=3,4$ are raised and lowered 
with $\delta^{ab}$, and the invariant antisymmetric tensors $\epsilon^{abc}$ (for 
$\cN=3$) and $\epsilon^{abcd}$ (for $\cN=4$) are normalized by $\epsilon^{123}=1$ and 
$\epsilon^{1234}=1$ respectively. The remaining conventions, including the explicit component expansion of superspace objects in $\theta$, follow our previous works \cite{AAN13,AJ22}.

\section{Analysis for $\epsilon$-invariants}
\label{app:epsilonrels}

\subsection{For $\cN=3$}
\label{app:epsilonrelsn3}

We list all candidate $\epsilon$-invariant structures at each $\lambda$-order, together with the linear relations among them. In each case the analysis yields precisely one independent parity-even and one independent parity-odd invariant (up to permutations of point labels), as quoted in Section~\ref{sec:epsilonn3}.

\begin{enumerate}[label=(\arabic*)]

	\item $\cO(\lambda_{1},\Theta^{3})$:
	\begin{align*}
	\text{indep even:\qquad}& \epsilon\,
	\Omega_{1}R_{1}\\
	\text{indep odd:\qquad}& \epsilon\,
	\Omega_{1}\omega_{1}
	\end{align*}
\
	\item $\cO(\lambda_{1}^{3},\Theta^{3})$:
	\begin{align*}
	\text{even:\qquad}& \epsilon\,
	R_{1}R_{1}R_{1},\ \epsilon\,R_{1}\omega_{1}\omega_{1}\\
	\text{odd:\qquad}& \epsilon\,
	\omega_{1}\omega_{1}\omega_{1},\ \epsilon\, R_{1}R_{1}\omega_{1}
	\end{align*}
	
	We have the relation:
	\begin{align}
	\epsilon\,
	\Omega_{1}R_{1}Q_{1}+
	\epsilon\,
	R_{1}R_{1}R_{1}-
	\epsilon\,
	R_{1}\omega_{1}\omega_{1}=0
	\end{align}
	and similarly for the odd ones.
	
	So,
	\begin{align*}
	\text{indep even:\qquad}& \epsilon\,
	R_{1}R_{1}R_{1}\\
	\text{indep odd:\qquad}& \epsilon\,
	R_{1}R_{1}\omega_{1}
	\end{align*}

	\item $\cO(\lambda_{1}^{2}\lambda_{2},\Theta^{3})$:
	\begin{align*}
	\text{even:\qquad}& \epsilon\,
	R_{1}R_{1}\sigma_{21},\ \epsilon\, \omega_{1}\omega_{1}\sigma_{21},\  \epsilon\, R_{1}\omega_{1}\pi_{21}
	\\
	\text{odd:\qquad}& \epsilon\,
	R_{1}R_{1}\pi_{21},\ \epsilon\, \omega_{1}\omega_{1}\pi_{21},\  \epsilon\, R_{1}\omega_{1}\sigma_{21}	
	\end{align*}
	
	We can also relate them to:
	\begin{align}
	\epsilon\,
	\Omega_{1}R_{1}P_{3}+
	\epsilon\,
	R_{1}\pi_{21}\omega_{1}-
	\epsilon\,
	R_{1}R_{1}\sigma_{21}&=0\\
	\epsilon\,
	\Omega_{2}R_{2}Q_{1}+
	\epsilon\,
	R_{1}R_{1}\sigma_{21}-
	\epsilon\,
	\omega_{1}\omega_{1}\sigma_{21}&=0
	\end{align}
	and similarly for the odd ones.
	
	Hence,
	\begin{align*}
	\text{indep even:\qquad}& \epsilon\,
	R_{1}R_{1}\sigma_{21}\\
	\text{indep odd:\qquad}& \epsilon\,
	R_{1}R_{1}\pi_{21}
	\end{align*}

	\item $\cO(\lambda_{1}\lambda_{2}\lambda_{3}	,\Theta^{3})$:
	\begin{align*}
	\text{even:\qquad}& \epsilon\,
	R_{1}\pi_{21}\pi_{31},\ \epsilon\, R_{1}\sigma_{21}\sigma_{31},\  \epsilon\, R_{2}\pi_{32}\pi_{12},\  \epsilon\, R_{3}\pi_{13}\pi_{23}
	\\
	\text{odd:\qquad}& \epsilon\,
	R_{1}\pi_{21}\sigma_{31},\ \epsilon\, R_{1}\sigma_{21}\pi_{31},\  \epsilon\, R_{2}\sigma_{32}\pi_{12},\  \epsilon\, \omega_{1}\pi_{21}\pi_{31}	
	\end{align*}
	
	We have relations:
	\begin{align}
	\epsilon\,
	\Omega_{1}R_{1}P_{1}-
	\epsilon\,
	\Omega_{3}R_{3}P_{3}+
	\epsilon\,
	R_{1}\pi_{21}\pi_{31}-
	\epsilon\,
	R_{3}\pi_{13}\pi_{23}&=0\\
	\epsilon\,
	\Omega_{2}R_{2}P_{2}-
	\epsilon\,
	\Omega_{3}R_{3}P_{3}+
	\epsilon\,
	R_{2}\pi_{32}\pi_{12}-
	\epsilon\,
	R_{3}\pi_{13}\pi_{23}&=0\\
	\epsilon\,
	\Omega_{3}R_{3}P_{3}+
	\epsilon\,
	R_{1}\sigma_{21}\sigma_{31}+
	\epsilon\,
	R_{3}\pi_{13}\pi_{23}=0
	\end{align}
	and similarly for the odd ones.
	
	Thus, 
	\begin{align*}
	\text{indep even:\qquad}& \epsilon\,
	R_{1}\sigma_{21}\sigma_{31}\\
	\text{indep odd:\qquad}& \epsilon\,
	\omega_{1}\pi_{21}\pi_{31}
	\end{align*}

\end{enumerate}

\subsection{For $\cN=4$}
\label{app:epsilonrelsn4}

Analogously to the $\cN=3$ case, we list all candidate structures at each $\lambda$-order, the linear relations reducing them, and the single independent parity-even and parity-odd $\epsilon$-invariant that survives at each order, as quoted in Section~\ref{sec:epsilonn4}.

\begin{enumerate}[label=(\arabic*)]

	\item $\cO(\text{no $\lambda$})$:
	\begin{align*}
	\text{even:\qquad}&
	\epsilon\,\Omega_{1}\Omega_{1},\\
	\text{odd:\qquad}&
	\text{none}.
	\end{align*}
	All permutations of $\epsilon\,\Omega_{1}\Omega_{1}$ coincide (since $\Omega_{i}^{ab}$ is symmetric in its point label for any fixed $ab$), so this single structure is the permutation-invariant $U'\equiv\epsilon\,\Omega_{1}\Omega_{1}$. It is not expressible as a polynomial in the $\cN=2$ fermionic invariants $R'$ and $T'$. The parity-odd candidate $\epsilon\,\Omega_{1}^{ab}\Omega_{1}^{ba}$ vanishes identically by antisymmetry of $\epsilon^{abcd}$ since it equals $-\epsilon\,\Omega_{1}\Omega_{1}$ up to reordering the R-symmetry indices; thus there is no independent parity-odd invariant at this order.
	\begin{align*}
	\text{indep even:\qquad}&
	\epsilon\,\Omega_{1}\Omega_{1},\\
	\text{indep odd:\qquad}&
	\text{none}.
	\end{align*}
	
	\item $\cO(\lambda_{1}^{2})$:
	\begin{align*}
	\text{even:\qquad}&
	\epsilon\,\Omega_{1}R_{1}R_{1},\ 
	\epsilon\,\Omega_{1}\omega_{1}\omega_{1},\\
	\text{odd:\qquad}&
	\epsilon\,\Omega_{1}R_{1}\omega_{1},
	\end{align*}
	
	We also have the relation
	\begin{align}
	\epsilon\,\Omega_{1}R_{1}R_{1}-\epsilon\,\Omega_{1}\omega_{1}\omega_{1}+\epsilon\,\Omega_{1}\Omega_{1}Q_{1}=0
	\end{align}
	Thus,
	\begin{align*}
	\text{indep even:\qquad}&
	\epsilon\,\Omega_{1}R_{1}R_{1},\\
	\text{indep odd:\qquad}&
	\epsilon\,\Omega_{1}R_{1}\omega_{1},
	\end{align*}
	
	\item $\cO(\lambda_{1}\lambda_{2})$:
	\begin{align*}
	\text{even:\qquad}&
	\epsilon\,\Omega_{1}R_{1}\sigma_{21},\
	\epsilon\,\Omega_{1}\omega_{1}\pi_{21},\\
	\text{odd:\qquad}&
	\epsilon\,\Omega_{1}\omega_{1}\sigma_{21},\
	\epsilon\,\Omega_{1}R_{1}\pi_{21},
	\end{align*}
	We have the relations
	\begin{align}
	\epsilon\,\Omega_{1}R_{1}\sigma_{21}+
	\epsilon\,\Omega_{1}\omega_{1}\pi_{21}+\epsilon\,\Omega_{1}\Omega_{1} P_{3}=0\\
	\epsilon\,\Omega_{1}\omega_{1}\sigma_{21}+
	\epsilon\,\Omega_{1}R_{1}\pi_{21}+\epsilon\,\Omega_{1}\Omega_{1} S_{3}=0
	\end{align}
	Thus,
	\begin{align*}
	\text{indep even:\qquad}&
	\epsilon\,\Omega_{1}R_{1}\sigma_{21},\\
	\text{indep odd:\qquad}&
	\epsilon\,\Omega_{1}R_{1}\pi_{21},
	\end{align*}
	
	\item $\cO(\lambda_{1}^{4})$:
	
	\begin{align*}
	\text{even:\qquad}&
	\epsilon\,R_{1}R_{1}R_{1}R_{1},\
	\epsilon\,R_{1}R_{1}\omega_{1}\omega_{1},\ 
	\epsilon\,\omega_{1}\omega_{1}\omega_{1}\omega_{1},\\
	\text{odd:\qquad}&
	\epsilon\,R_{1}R_{1}R_{1}\omega_{1},\
	\epsilon\,R_{1}\omega_{1}\omega_{1}\omega_{1},\ 
	\end{align*}
	We have the relations
	\begin{align}
	\epsilon\,R_{1}R_{1}R_{1}R_{1}-
	\epsilon\,R_{1}R_{1}\omega_{1}\omega_{1}+\epsilon\,\Omega_{1}R_{1}R_{1}Q_{1}=0\\
	\epsilon\,R_{1}R_{1}R_{1}R_{1}-
	\epsilon\,\omega_{1}\omega_{1}\omega_{1}\omega_{1}+2\epsilon\,\Omega_{1}R_{1}R_{1}Q_{1}+\epsilon\,\Omega_{1}\Omega_{1}Q_{1}^{2}&=0
	\end{align}
	
	Thus,
	\begin{align*}
	\text{indep even:\qquad}&
	\epsilon\,R_{1}R_{1}R_{1}R_{1},\\
	\text{indep odd:\qquad}&
	\epsilon\,R_{1}R_{1}R_{1}\omega_{1},
	\end{align*}

	\item $\cO(\lambda_{1}^{3}\lambda_{2})$:
	\begin{align*}
	\text{even:\qquad}&
	\epsilon\,R_{1}R_{1}R_{1}\sigma_{21},\ 
\epsilon\,R_{1}\omega_{1}R_{1}\sigma_{21},\ 
\epsilon\,R_{1}R_{1}\omega_{1}\pi_{21},\ 
\epsilon\,\omega_{1}\omega_{1}\omega_{1}\pi_{21},\\
	\text{odd:\qquad}&
	\epsilon\,R_{1}R_{1}R_{1}\pi_{21},\ 
\epsilon\,R_{1}\omega_{1}R_{1}\pi_{21},\ 
\epsilon\,R_{1}R_{1}\omega_{1}\sigma_{21},\ 
\epsilon\,\omega_{1}\omega_{1}\omega_{1}\sigma_{21}
	\end{align*}
	We have the relations
	\begin{gather}
	\epsilon\,R_{1}R_{1}R_{1}\sigma_{21}
- \epsilon\,R_{1}\omega_{1}\omega_{1}\sigma_{21}
+ \epsilon\,\Omega_{1}R_{1}\sigma_{21}Q_{1}=0,\\
\epsilon\,R_{1}R_{1}R_{1}\sigma_{21}
+ \epsilon\,R_{1}R_{1}\omega_{1}\pi_{21}
+ \epsilon\,\Omega_{1}R_{1}R_{1}P_{3}=0,\\
\epsilon\,R_{1}R_{1}R_{1}\sigma_{21}
+ \epsilon\,\omega_{1}\omega_{1}\omega_{1}\pi_{21}
+ \epsilon\,\Omega_{1}\Omega_{1}Q_{1}P_{3}
+ \epsilon\,\Omega_{1}R_{1}R_{1}P_{3}
+ \epsilon\,\Omega_{1}R_{1}\sigma_{21}Q_{1}=0,
	\end{gather}

	Thus,
	\begin{align*}
	\text{indep even:\qquad}&
	\epsilon\,R_{1}R_{1}R_{1}\sigma_{21},\\
	\text{indep odd:\qquad}&
	\epsilon\,R_{1}R_{1}R_{1}\pi_{21},
	\end{align*}

	\item $\cO(\lambda_{1}^{2}\lambda_{2}^{2})$:

	At this $\lambda$-order the even and odd sectors each contain several candidate structures, formed from pairs of degree-$(\lambda_{1})$ covariants $\{R_{1},\omega_{1}\}$ and pairs of degree-$(\lambda_{2})$ covariants $\{R_{2},\omega_{2},\pi_{21},\sigma_{21}\}$ (and the mixed $\{R_{2},\omega_{2}\}\leftrightarrow\{\pi_{12},\sigma_{12}\}$ counterparts). Linear relations reduce each sector, and the requirement that the surviving structures have a definite transformation property under exchange $1\leftrightarrow 2$ imposes a further constraint. One finds a single independent even and a single independent odd invariant:
	\begin{align*}
	\text{indep even:\qquad}&
	\epsilon\,\omega_{1}\omega_{1}\pi_{21}\pi_{21},\\
	\text{indep odd:\qquad}&
	\epsilon\,R_{1}R_{1}\pi_{21}\sigma_{21}-\epsilon\,R_{2}R_{2}\pi_{12}\sigma_{12}.
	\end{align*}
	The antisymmetric combination in the odd sector is dictated by permutation symmetry: $U_{220}$ must change sign under $1\leftrightarrow2$ (see Section~\ref{sec:permsym}), which is satisfied by the difference above but not by either term individually.

	\item $\cO(\lambda_{1}^{2}\lambda_{2}\lambda_{3})$:

	With three distinct point labels appearing asymmetrically (degree 2 in $\lambda_{1}$, degree 1 each in $\lambda_{2}$ and $\lambda_{3}$), the candidate structures include products of two $\{R_{1},\omega_{1}\}$ covariants with one degree-$(\lambda_{2})$ and one degree-$(\lambda_{3})$ covariant. After reducing by linear relations one finds:
	\begin{align*}
	\text{indep even:\qquad}&
	\epsilon\,R_{1}R_{1}\pi_{21}\pi_{31},\\
	\text{indep odd:\qquad}&
	\epsilon\,R_{1}\omega_{1}\pi_{21}\pi_{31}.
	\end{align*}

\end{enumerate}

\bibliographystyle{JHEP}
\bibliography{n34scft}

@article{AAN13,
    author        = "Nizami, Amin A. and Sharma, Tarun and Umesh, V.",
    title         = "{Superspace formulation and correlation functions of 3d superconformal field theories}",
    eprint        = "1308.4778",
    archivePrefix = "arXiv",
    primaryClass  = "hep-th",
    doi           = "10.1007/JHEP07(2014)022",
    journal       = "JHEP",
    volume        = "07",
    pages         = "022",
    year          = "2014"
}

@article{AJ22,
    author        = "Jain, Aditya and Nizami, Amin A.",
    title         = "{Superconformal invariants and spinning correlators in 3d $\mathcal{N}=2$ SCFTs}",
    eprint        = "2205.11157",
    archivePrefix = "arXiv",
    primaryClass  = "hep-th",
    doi           = "10.1140/epjc/s10052-022-11016-2",
    journal       = "Eur. Phys. J. C",
    volume        = "82",
    pages         = "1065",
    year          = "2022"
}

@article{AJ24,
    author        = "Jain, Aditya and Nizami, Amin A.",
    title         = "{Superspace invariants and correlators in 4d $\mathcal{N}=1$ superconformal field theories}",
    eprint        = "2411.01903",
    archivePrefix = "arXiv",
    primaryClass  = "hep-th",
    doi           = "10.1007/JHEP02(2025)037",
    journal       = "JHEP",
    volume        = "02",
    pages         = "037",
    year          = "2025"
}

@article{Park97,
    author        = "Park, Jeong-Hyuck",
    title         = "{$\mathcal{N}=1$ superconformal symmetry in four-dimensional quantum field theory}",
    eprint        = "hep-th/9703191",
    archivePrefix = "arXiv",
    doi           = "10.1142/S0217751X98000695",
    journal       = "Int. J. Mod. Phys. A",
    volume        = "13",
    pages         = "1743--1772",
    year          = "1998"
}

@article{Park98,
    author        = "Park, Jeong-Hyuck",
    title         = "{Superconformal symmetry and correlation functions}",
    eprint        = "hep-th/9903230",
    archivePrefix = "arXiv",
    doi           = "10.1016/S0550-3213(99)00432-0",
    journal       = "Nucl. Phys. B",
    volume        = "559",
    pages         = "455--501",
    year          = "1999"
}

@article{Park99,
    author        = "Park, Jeong-Hyuck",
    title         = "{Superconformal symmetry in three dimensions}",
    eprint        = "hep-th/9910199",
    archivePrefix = "arXiv",
    doi           = "10.1063/1.1290382",
    journal       = "J. Math. Phys.",
    volume        = "41",
    pages         = "7129--7161",
    year          = "2000"
}

@article{Osborn94,
    author        = "Osborn, H. and Petkou, A. C.",
    title         = "{Implications of conformal invariance in field theories for general dimensions}",
    eprint        = "hep-th/9307010",
    archivePrefix = "arXiv",
    doi           = "10.1006/aphy.1994.1045",
    journal       = "Ann. Phys.",
    volume        = "231",
    pages         = "52--121",
    year          = "1994"
}

@article{Osborn98,
    author        = "Osborn, H.",
    title         = "{$\mathcal{N}=1$ superconformal symmetry in four dimensional quantum field theory}",
    eprint        = "hep-th/9808041",
    archivePrefix = "arXiv",
    doi           = "10.1006/aphy.1998.5826",
    journal       = "Ann. Phys.",
    volume        = "272",
    pages         = "243--294",
    year          = "1999"
}

@article{Erd97,
    author        = "Erdmenger, J. and Osborn, H.",
    title         = "{Conserved currents and the energy-momentum tensor in conformally invariant theories for general dimensions}",
    eprint        = "hep-th/9605009",
    archivePrefix = "arXiv",
    doi           = "10.1016/S0550-3213(96)00545-7",
    journal       = "Nucl. Phys. B",
    volume        = "483",
    pages         = "431--474",
    year          = "1997"
}

@article{Dolan01,
    author        = "Dolan, F. A. and Osborn, H.",
    title         = "{Correlation functions of conserved currents in $\mathcal{N}=1$ superconformal theories}",
    eprint        = "hep-th/0111228",
    archivePrefix = "arXiv",
    doi           = "10.1006/aphy.2002.6299",
    journal       = "Ann. Phys.",
    volume        = "307",
    pages         = "41--89",
    year          = "2003"
}

@article{GPY11,
    author        = "Giombi, Simone and Prakash, Shiroman and Yin, Xi",
    title         = "{A Note on CFT Correlators of Gauge Invariant Operators}",
    eprint        = "1104.4317",
    archivePrefix = "arXiv",
    primaryClass  = "hep-th",
    doi           = "10.1007/JHEP07(2013)105",
    journal       = "JHEP",
    volume        = "07",
    pages         = "105",
    year          = "2013"
}

@article{BKS15,
    author        = "Buchbinder, Evgeny I. and Kuzenko, Sergei M. and Samsonov, Igor B.",
    title         = "{Superconformal field theory in three dimensions: Correlation functions of conserved currents}",
    eprint        = "1503.04961",
    archivePrefix = "arXiv",
    primaryClass  = "hep-th",
    doi           = "10.1007/JHEP06(2015)138",
    journal       = "JHEP",
    volume        = "06",
    pages         = "138",
    year          = "2015"
}

@article{BKS15b,
    author        = "Buchbinder, Evgeny I. and Kuzenko, Sergei M. and Samsonov, Igor B.",
    title         = "{Implications of $\mathcal{N}=4$ superconformal symmetry in three spacetime dimensions}",
    eprint        = "1507.00221",
    archivePrefix = "arXiv",
    primaryClass  = "hep-th",
    doi           = "10.1007/JHEP08(2015)125",
    journal       = "JHEP",
    volume        = "08",
    pages         = "125",
    year          = "2015"
}

@article{KS16,
    author        = "Kuzenko, Sergei M. and Samsonov, Igor B.",
    title         = "{Implications of $\mathcal{N}=5,6$ superconformal symmetry in three spacetime dimensions}",
    eprint        = "1605.08208",
    archivePrefix = "arXiv",
    primaryClass  = "hep-th",
    doi           = "10.1007/JHEP08(2016)084",
    journal       = "JHEP",
    volume        = "08",
    pages         = "084",
    year          = "2016"
}

@article{BS21,
    author        = "Buchbinder, Evgeny I. and Stone, Benjamin J.",
    title         = "{Mixed three-point functions of conserved currents in three-dimensional superconformal field theory}",
    eprint        = "2102.04827",
    archivePrefix = "arXiv",
    primaryClass  = "hep-th",
    doi           = "10.1103/PhysRevD.103.086023",
    journal       = "Phys. Rev. D",
    volume        = "103",
    pages         = "086023",
    year          = "2021"
}

@article{BS21b,
    author        = "Buchbinder, Evgeny I. and Stone, Benjamin J.",
    title         = "{Three-point functions of a superspin-2 current multiplet in 3D, $\mathcal{N}=1$ superconformal theory}",
    eprint        = "2108.01865",
    archivePrefix = "arXiv",
    primaryClass  = "hep-th",
    doi           = "10.1103/PhysRevD.104.106004",
    journal       = "Phys. Rev. D",
    volume        = "104",
    pages         = "106004",
    year          = "2021"
}

@article{BS23a,
    author        = "Buchbinder, Evgeny I. and Stone, Benjamin J.",
    title         = "{Three-point functions of conserved supercurrents in 3D $\mathcal{N}=1$ SCFT: general formalism for arbitrary superspins}",
    eprint        = "2302.00593",
    archivePrefix = "arXiv",
    primaryClass  = "hep-th",
    doi           = "10.1103/PhysRevD.107.106001",
    journal       = "Phys. Rev. D",
    volume        = "107",
    pages         = "106001",
    year          = "2023"
}

@article{BS23b,
    author        = "Buchbinder, Evgeny I. and Stone, Benjamin J.",
    title         = "{Grassmann-odd three-point functions of conserved supercurrents in 3D $\mathcal{N}=1$ SCFT}",
    eprint        = "2305.02233",
    archivePrefix = "arXiv",
    primaryClass  = "hep-th",
    doi           = "10.1103/PhysRevD.108.046001",
    journal       = "Phys. Rev. D",
    volume        = "108",
    pages         = "046001",
    year          = "2023"
}

@article{MZ1,
    author        = "Maldacena, Juan and Zhiboedov, Alexander",
    title         = "{Constraining Conformal Field Theories With A Higher Spin Symmetry}",
    eprint        = "1112.1016",
    archivePrefix = "arXiv",
    primaryClass  = "hep-th",
    doi           = "10.1088/1751-8113/46/21/214011",
    journal       = "J. Phys. A",
    volume        = "46",
    pages         = "214011",
    year          = "2013"
}

@article{MZ2,
    author        = "Maldacena, Juan and Zhiboedov, Alexander",
    title         = "{Constraining conformal field theories with a slightly broken higher spin symmetry}",
    eprint        = "1204.1698",
    archivePrefix = "arXiv",
    primaryClass  = "hep-th",
    doi           = "10.1088/0264-9381/30/10/104003",
    journal       = "Class. Quant. Grav.",
    volume        = "30",
    pages         = "104003",
    year          = "2013"
}

\end{document}